\documentclass[a4paper,
11pt,unpublished]{quantumarticle}
\pdfoutput=1

\usepackage[utf8]{inputenc}
\usepackage[english]{babel}
\usepackage{amsthm}
\usepackage{amsmath}
\usepackage{amssymb}
\usepackage{stmaryrd}
\usepackage{mathtools}
\usepackage{pifont}
\usepackage{physics}
\usepackage{hyperref}
\usepackage{ bbold }
\usepackage{ebproof}
\usepackage{listings}
\usepackage{thm-restate}
\usepackage{enumitem}
\usepackage{xfrac}

\usepackage{tikz}
\usetikzlibrary{quantikz2}
\usetikzlibrary{patterns}

\newcommand{\compgroup}[2]{\gategroup[#1,steps=#2,style={dashed,fill = gray!0, inner sep=0pt},background]{}}

\newcommand{\maxcolor}[1]{{\color{blue} #1}}

\tikzset{
mystyle/.style ={preaction={fill, white}, pattern =north west lines, pattern color=black}
}
\tikzset{
mystyle2/.style ={preaction={fill, white},pattern =
crosshatch dots, pattern color=black}
}

\newtheorem{definition}{Definition}

\usepackage{graphicx} 

\usepackage{cellspace} 
\usepackage{xcolor}

\definecolor{procedures}{RGB}{202,82,213}

\definecolor{commands}{RGB}{9,10,184}

\usepackage{amsfonts}

\usepackage{pgfplots}
\usepackage{pgfplotstable}
\pgfplotsset{compat=1.18}

\newcommand{\pfoqcompiler}{\texttt{pfoqcompiler}}

\newcommand{\ia}{\ensuremath{\mathrm{i}}}

\newcommand{\tb}[1]{\text{\textnormal{\textbf{#1}}}}
\newcommand{\tbt}[1]{\text{\color{commands}\textnormal{\texttt{#1}}}}
\newcommand{\pn}[1]{\text{\color{procedures}\textnormal{\texttt{#1}}}}

\newcommand{\proc}{\pn{proc}}

\newcommand{\call}{\tbt{call }}
\renewcommand{\ket}[1]{\ensuremath{\left| #1\right\rangle}}

\newcommand{\ST}{\textnormal{\texttt{S}}}

\newcommand{\PR}{\textnormal{\texttt{P}}}
\newcommand{\QFT}{\ensuremath{\texttt{\color{gray}QFT}}}

\newcommand{\asg}{{\ \textnormal{\texttt{\textasteriskcentered =}}\ }}

\newcommand{\width}{\textnormal{width}}

\newcommand{\foq}{\textnormal{\textsc{foq}}}
\newcommand{\wf}{\textnormal{\textsc{wf}}}
\newcommand{\wi}{\textnormal{\textsc{width}}}
\newcommand{\half}{\textnormal{\textsc{half}}}

\newcommand{\pfoq}{\textnormal{\textsc{pfoq}}}
\newcommand{\lfoq}{\textnormal{\textsc{lfoq}}}
\newcommand{\unif}{\textnormal{\textsc{unif}}}

\usepackage{listings}

\definecolor{procedures}{RGB}{202,82,213}

\definecolor{commands}{RGB}{9,10,184}

\newenvironment{PFOQprogram}{\ttfamily \begin{tabular}{l} }{\end{tabular}}

\title{pfoqcompiler: Efficient Compilation of Quantum Control in Recursive First-Order Quantum Programs} 

\author{Mário Silva}
\email{mariosilva@lmf.cnrs.fr}
\affiliation{Universit\'{e} Paris-Saclay, CNRS, Inria, LMF, 91190 Gif-sur-Yvette, France}
\orcid{0000-0002-9886-8400}
\author{Alexandre Guernut}
\affiliation{Université de Lorraine, CNRS, Inria, LORIA, 54000 Nancy, France}
\orcid{0009-0003-0152-632X}
\author{Marco Lewis}
\affiliation{Universit\'{e} Paris-Saclay, CNRS, Inria, LMF, 91190 Gif-sur-Yvette, France}
\orcid{0000-0002-4893-7658}

\begin{document}

\maketitle
 
\begin{abstract}
Recursive quantum control allows for simple and concise descriptions of algorithms. However, a naive circuit compilation sequentializes the different branches of the quantum if-statement and results in an exponential increase in circuit depth. Existing techniques can parallelize these branches, but do so at the cost of an exponential number of ancillary qubits.

We present \pfoqcompiler{}, a compiler targeting Qiskit for the first-order quantum programming language \foq{} that exploits program structure to merge compatible recursive procedure calls across quantum branches. We introduce weaker syntactically-checkable conditions for well-foundedness, recursive halving, and uniformity, while preserving previously demonstrated complexity characterizations. For weakly well-founded bounded-width programs, the compiler produces polynomial-size circuits; weakly recursively halving programs additionally yield polylogarithmic depth, while weak uniformity removes the asymptotic overhead of branch sequentialization.
\end{abstract}

\section{Introduction}
\label{s:introduction}

\subsection{The complexity of quantum control}

There has been an interest in the development of programming languages that allow for \emph{quantum control}, that is, the ability to guide the flow of computation according to the state of a qubit without performing a measurement, allowing for the execution of program instructions in quantum superposition.

From the point of view of a quantum system, this corresponds to a unitary of the form:
\begin{equation}
\ketbra{0}{0}\otimes U_0 + \ketbra{1}{1}\otimes U_1
\end{equation} 
where $U_0$ and $U_1$ are unitary operators to be applied in the branches where the \emph{control qubit} is in state $\ket{0}$ or $\ket{1}$, respectively. This can be easily generalized to $k$ control qubits with $2^k$ different unitaries $U_i$, $0\leq i \leq 2^k-1$.

A straightforward circuit implementing quantum control consists of applying each $\ketbra{i}\otimes U_i$ in sequence, resulting in a circuit with time complexity given by the sum of the time complexities of each $U_i$. This approach has an important consequence:  the complexity of quantum branching is not the maximum of each branch (as in the deterministic or probabilistic cases) but rather their \emph{sum}. This problem has been termed \emph{branch sequentialization} in~\cite{TOWER}. 

While in many scenarios branch sequentialization does not increase a program's time complexity, it can have a large impact if the quantum control statement is used \emph{recursively}, leading to circuits of exponential size even for relatively simple programs.

\subsection{Taming quantum recursion}

Many quantum programming languages include a quantum control instruction which can be used recursively. Such examples include 
Quipper
 ~\cite{QUIPPER}, Silq~\cite{SILQ,SILQ_HPDC25}, Tower~\cite{TOWER}, Qunity~\cite{QUNITY,QUNITY_compile}, Qrisp~\cite{QRISP}, and RQC$^{++}$~\cite{ZY25}. Almost all of these languages allow for compilation into quantum circuits, and a few of them have a focus on the efficiency of their implementation of quantum control.

More notably, in~\cite{ZY25}, the authors investigate techniques for automatically parallelizing quantum branches, with the express purpose of studying how quantum recursive programs can be efficiently
implemented. Based on the techniques inspired by QRAM and quantum multiplexing~\cite{ZLY22}, the authors provide a compilation strategy that automatically parallelizes recursive quantum procedure calls, which results in an exponential speedup in circuit depth.

Qrisp~\cite{QRISP} also allows the user to parallelize instructions in quantum-controlled statements, alongside other compilation methods. Of note, the compilation techniques of~\cite{ZLY22,QRISP} require an exponential number of ancillary qubits in the general case.

These approaches are essentially limited by the lack of knowledge about the structure of the circuit describing each $U_i$. If these subcircuits are treated as black boxes, then the final circuit \emph{size} will necessarily be given by the sum of the complexities of the black boxes.

\subsection{Leveraging program structure: first-order quantum programs}

Another approach is to restrict the language such that its structure can be exploited to reduce the asymptotic circuit complexity. Recent compilation techniques stemming from the field of implicit computational complexity  simplify recursive procedure calls that occur repeatedly in the circuit's compilation, with a method of \emph{anchoring and merging}. The \foq{} (First-Order Quantum) programming language introduced in~\cite{HPS23} and further developed in~\cite{FHPS25,HPS25,hainry2025polytime}, avoids the exponential blow-up of recursive quantum branching both in circuit width (i.e. number of qubits and ancillae) and depth.

We recall here the illustrative example appearing in~\cite{FHPS25} of a recursive procedure \pn{pairs} that detects whether an input string is written with pairs of values 00 and 11. Put differently, we wish to write a program that describes a unitary $U$ such that, for $\bar{x}\in\{0,1\}^*,\, y \in\{0,1\}$,
\begin{equation}
U \ket{\bar{x}y} = \ket{\bar{x}(y\oplus b)},
\end{equation}
where $b\in\{0,1\}$ and
\begin{equation}
b=1 \Leftrightarrow \bar{x}\in(00|11)^\ast. 
\end{equation} 

A program describing such a unitary operator lends itself easily to a quantum recursive function, as described in Figure~\ref{fig:pairs}.

\begin{figure}
\begin{minipage}{0.45\columnwidth}
\scalebox{0.85}{
\begin{PFOQprogram}
\tbt{decl }\pn{pairs}(a)\{\\
\quad \tbt{if }(|a|>1)\tbt{ then }\{ {\color{gray} \# recursive case}\\
\qquad  \tbt{qcase }(a[0],a[1])\tbt{ of }\{ \\
\qquad\quad  00 -> \tbt{call }\pn{pairs}(a-[0,1]);,\\
 \qquad\quad 01 -> \tbt{skip};,\\
 \qquad\quad 10 -> \tbt{skip};,\\
\qquad\quad  11  -> \tbt{call }\pn{pairs}(a-[0,1]);\ \}\\
\quad  \} \tbt{else} \{ a[0] *= NOT; \}\ \} {\color{gray} \# base case}
\end{PFOQprogram}}
\end{minipage}
\caption{\foq{} procedure detecting strings in $(00|11)^\ast$.}
\label{fig:pairs}
\end{figure}

The \pn{pairs} procedure takes as input a qubit list variable \texttt{a} and, if the length of the list is larger than one, it performs a quantum control statement where \pn{pairs} calls itself recursively if the first two qubits are in the same state, and performs a no-op instruction otherwise. By language construction, \pn{pairs} terminates whenever its input is empty.

For an input list \texttt{a} of length $n$, the base case is only reached if the state of the first $n-1$ qubits is described by the regular language $(00|11)^*$, in which case a $NOT$ gate is applied to the last qubit, as described in the definition of $U$. 

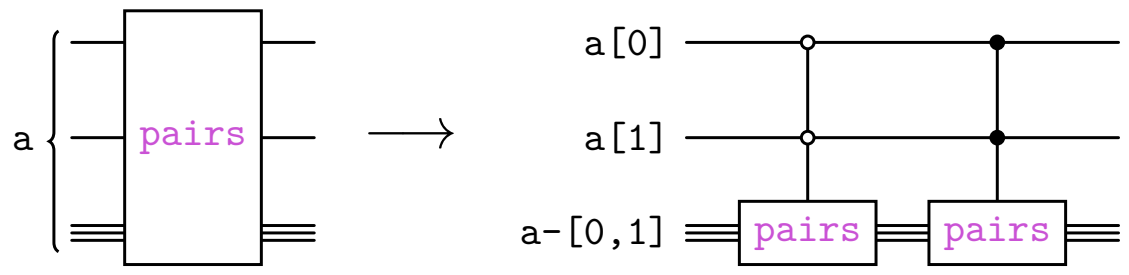
\begin{figure}[h!]
\[\scalebox{0.7}{
\begin{quantikz}[wire types = {q,q,b},classical gap=0.07cm, row sep = {9mm, between origins}]
\lstick[3]{\texttt{a}} & \gate[3]{\pn{pairs}} & \\
& &\\
& &
\end{quantikz}}\ 
\longrightarrow\ 
\scalebox{0.7}{
\begin{quantikz}[wire types = {q,q,b},classical gap=0.07cm,row sep = {9mm, between origins}]
\lstick{\texttt{a[0]}} & \octrl{1} & \ctrl{1} & \\
\lstick{\texttt{a[1]}} & \octrl{1} & \ctrl{1} & \\
\lstick{\texttt{a-[0,1]}} &  \gate{\pn{pairs}}& \gate{\pn{pairs}}&
\end{quantikz}}\]
\caption{The sequential strategy.}
\label{fig:sequential-strategy}
\end{figure}

The combined use of recursion and a quantum if-statement allows for the compact definition of \pn{pairs}, and we now turn to its compilation. A straightforward implementation of a quantum if-statement would give us the \emph{sequential} strategy for quantum control described in Figure~\ref{fig:sequential-strategy}.

Given the doubling of instances of \pn{pairs} in sequence at every iteration, the sequential strategy results in a circuit of exponential depth.

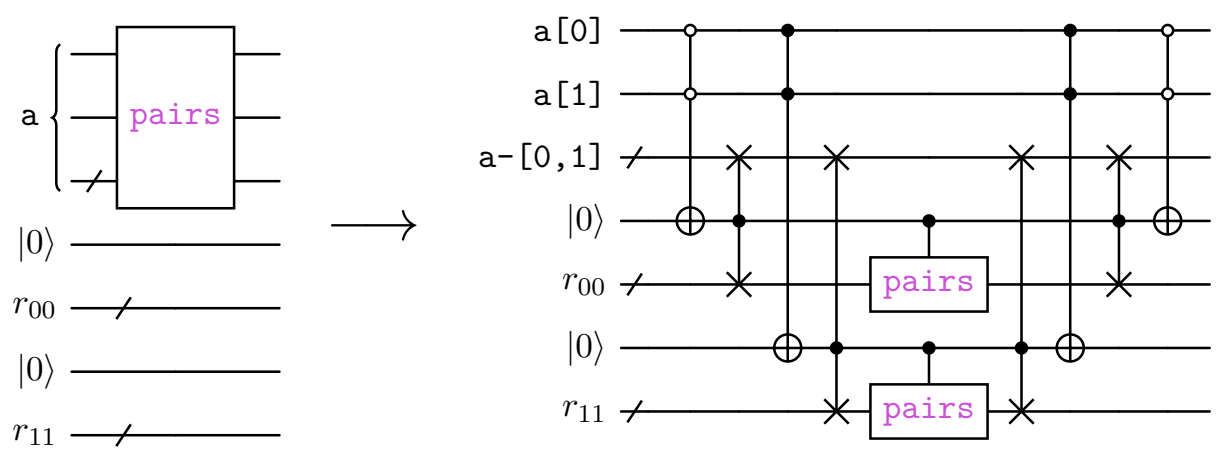
\begin{figure}[h!]
\[
\scalebox{0.6}{
\begin{quantikz}[classical gap=0.07cm, row sep = {7mm, between origins}]
\lstick[3]{\texttt{a}} & \gate[3]{\pn{pairs}} & \\
& &\\
& \qwbundle{} &\\
\lstick{$\ket{0}$} & & \\
\lstick{$r_{00}$} &\qwbundle{} & \\
\lstick{$\ket{0}$} & & \\
\lstick{$r_{11}$} &\qwbundle{} & 
\end{quantikz}}\ 
\longrightarrow\ 
\scalebox{0.6}{
\begin{quantikz}[classical gap=0.07cm, column sep = 3mm, row sep = {7mm, between origins}]
\lstick{\texttt{a[0]}}  & & \octrl{1} & &\ctrl{1} & & &&\ctrl{1} & &\octrl{1} &\\
\lstick{\texttt{a[1]}} & &\octrl{2} & &\ctrl{4} & & & & \ctrl{4} & &\octrl{2} &\\
\lstick{\texttt{a-[0,1]}} & \qwbundle{} & &\swap{2} & & \swap{4} &  & \swap{4} & &\swap{2} & &\\
\lstick{$\ket{0}$} & &\targ{} & \ctrl{0}& & & \ctrl{1} & & & \ctrl{0} &\targ{} &\\
\lstick{$r_{00}$} &\qwbundle{} & &\targX{} & & &\gate{\pn{pairs}} & & &\targX{} & &\\
\lstick{$\ket{0}$} & & & & \targ{} & \ctrl{0} & \ctrl{1} &\ctrl{0}& \targ{} && &\\
\lstick{$r_{11}$} &\qwbundle{} & & & &\targX{} &\gate{\pn{pairs}} & \targX{} & & & &
\end{quantikz}}
\]
\caption{The parallel strategy.}
\label{fig:parallel-strategy}
\end{figure}

We can consider an alternative strategy. Using fresh registers $r_{00}$ and $r_{11}$ for each of the non-trivial branches of the quantum if-statement, one can implement each branch in parallel (Figure~\ref{fig:parallel-strategy}).

This \emph{parallel strategy} employs controlled swapping of registers which consists of controlled permutations and can be performed in logarithmic depth and linear size~\cite{ZLY22,QRISP}. The parallel strategy prevents the recursive doubling in depth of the sequential strategy, but requires an exponential number of operations and ancillae.

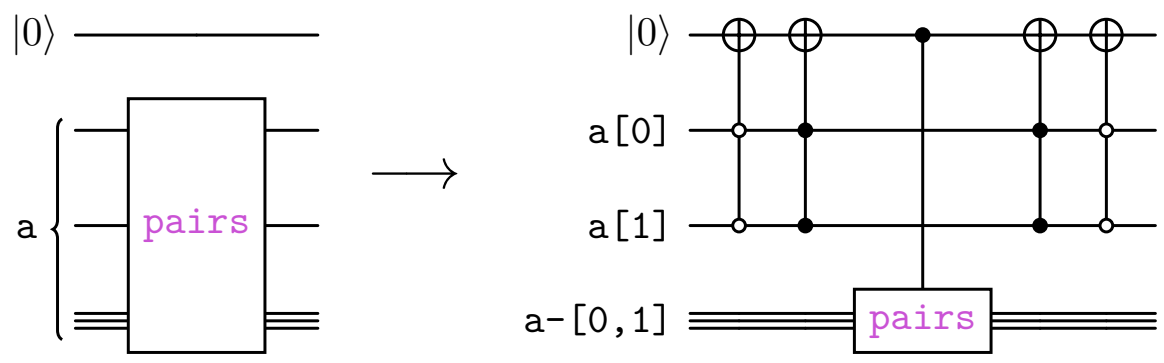
\begin{figure}[h!]
\[
\scalebox{0.7}{
\begin{quantikz}[wire types = {q,q,q,b},classical gap=0.07cm, row sep = {9mm, between origins}]
\lstick{$\ket{0}$} & & \\
\lstick[3]{\texttt{a}} & \gate[3]{\pn{pairs}} & \\
& &\\
& &
\end{quantikz}}\ 
\longrightarrow\ 
\scalebox{0.7}{
\begin{quantikz}[wire types = {q,q,q,b},classical gap=0.07cm,row sep = {9mm, between origins}, column sep = 3mm]
\lstick{$\ket{0}$} & \targ{} & \targ{}  & \ctrl{3} & \targ{} & \targ{} & \\
\lstick{\texttt{a[0]}} & \octrl{-1} & \ctrl{-1} &  & \ctrl{-1} & \octrl{-1} &\\
\lstick{\texttt{a[1]}} & \octrl{-1} & \ctrl{-1} & & \ctrl{-1} & \octrl{-1} & \\
\lstick{\texttt{a-[0,1]}} &   &  & \gate{\pn{pairs}} & & & 
\end{quantikz}}
\]
\caption{The merging strategy.}
\label{fig:merging-strategy}
\end{figure}

If we now consider the structure of the \pn{pairs} program, a straightforward simplification is available: since the two non-trivial branches are identical, their quantum controls can be combined in the state of a single ancilla, giving us the \emph{merging} strategy in Figure~\ref{fig:merging-strategy}.

The merging strategy avoids the doubling of procedure calls and achieves a circuit of linear complexity (depth, size and total number of qubits). Notably, linear complexity is precisely what would be expected if the time complexity of quantum branching were given by the maximum complexities instead of the sum.

Given the simplicity of the \pn{pairs} procedure, this compilation strategy could be found manually. The objective of \texttt{pfoqcompiler}, however, is to generalize this approach to a large set of programs, for which a simplification as described above is not readily apparent. For instance, we show that a small change to \pn{pairs} makes the merging strategy highly non-trivial (Section~\ref{s:compilation-example}).

\subsection{Implicit computational complexity}

The compilation method of anchoring and merging will allow us to reduce the complexity of quantum recursion such that the complexity cost of the quantum if-statement can be taken as the maximum of the complexities of each of the branches, as in the classical and probabilistic cases. This can be achieved not over the entirety of all first-order quantum programs, but rather over a subset, which will have enough structure such that the circuit optimization is possible.

Since we are mainly interested in obtaining efficient circuits for programs which are themselves efficient, we can restrict our approach to polynomial-time programs. In~\cite{HPS23}, the authors provide restrictions for the \foq{} programming language that result in fragments that are sound and complete for quantum polynomial-time~\cite{HPS23, HPS25} and polylogarithmic-time programs~\cite{FHPS25}.

\foq{}'s polynomial-time fragment, denoted \pfoq{}, requires two program restrictions: \emph{well-foundedness} and \emph{bounded width}. Well-foundedness ensures termination by requiring that mutually recursive procedure calls reduce the number of available resources -- in this case, input qubits. We recall the \emph{strong} form of this requirement introduced in~\cite{HPS23} in Section~\ref{sec:restrictions} and introduce a \emph{weaker} version of well-foundedness which we prove to be extensionally equivalent and therefore sound for termination.

For polylogarithmic-time programs, which can be compiled into circuits of polylogarithmic depth, a stronger restriction than well-foundedness is considered, denoted \emph{recursively halving}, which requires that some input qubit list is reduced by half for every recursive call. This clearly subsumes well-foundedness, and we also give a weak version of this constraint. 

The restriction of bounded width requires that procedures contain at most one recursive call per branch of computation, meaning that multiple such calls may appear in different branches of a classical or quantum if-statement, but may not be performed in sequence. This property is crucial for the validity of the compilation strategy but is also justified from the point-of-view of complexity, as programs with unbounded width typically run in exponential time.

Finally, we address a strict subset of \pfoq{} programs that can be compiled such that the cost of quantum branching is precisely the maximum of all branches. This subset satisfies the property of \emph{uniformity}, which limits the ways in which procedure calls can differ and how qubits can be removed from the input.

This uniform subset, while still sound and complete for quantum polynomial time~\cite{HPS25}, is quite restrictive in terms of program design and therefore we give a weaker version of uniformity that satisfies the same properties.

\subsection{Contributions}

The main contribution of this work is a software tool, the \pfoqcompiler{}~\cite{pfoqcompiler_zenodo}, which:
\begin{itemize}
\item Provides the first implementation of the \foq{} programming language, combining recursion with quantum and classical conditional statements.
\item Includes static analysis tools that efficiently verify the syntactic restrictions on \foq{} programs related to different implicit complexity characterizations, including polynomial and polylogarithmic-time restrictions.
\end{itemize}

The \pfoqcompiler{} is also built on new theoretical results, given in Section~\ref{sec:restrictions}, which render the compiler (and the \foq{} language) more practically applicable. The theoretical results are the following:
\begin{itemize}
\item We define weaker versions of qubit removal restrictions, namely \emph{weak} well-foundedness and \emph{weak} recursive halving, given in Definitions~\ref{def:weak-wf} and~\ref{def:weak-rec-halv}, respectively. These new restrictions impose qubit list reduction only once per loop of recursive calls, and we show that they are \emph{extensionally equivalent} to their strong versions, i.e. that a weakly well-founded (or recursively halving) program can be simulated by its strong version (Theorem~\ref{thm:wfstar}) and so the restrictions preserve the same complexity properties.
\item For the third relaxation, we extend the notion of \emph{uniformity} in a program to include \emph{front-reducing} and \emph{back-reducing} programs (qubits are always removed from the beginning of the list, or the end, respectively) and prove that uniform programs allow for an optimal circuit complexity bound in Theorem~\ref{thm:weak-unif-circ}.
\item We introduce the first compiler for the \foq{} programming language, translating programs into circuits in the Qiskit language. The circuits can then be directly translated into OpenQASM or further improved using Qiskit optimization tools.
\item We experimentally verify the complexity bounds of circuits corresponding to different \foq{} fragments, and compare these results with other programming languages that allow for quantum recursive branching. 
\end{itemize}

In this work, we bring the ideas of~\cite{HPS23,HPS25,FHPS25} to practical fruition not only by providing a compiler that implements the techniques described in these papers, but also by relaxing some of the restrictions placed on programs to which the circuit bounds apply, making the compilation techniques more generally applicable.

We also implement efficient static verification of all program restrictions described in this work, allowing for quick feedback to the programmer on the complexity of resulting circuits.

\subsection{Related work}

\paragraph*{Implicit computational complexity.}
The resource use of syntactically-restricted languages is the field of study of \emph{implicit computational complexity} (ICC)~\cite{DL11,P20}. Languages characterizing efficient quantum computation have been given in the context of function algebras~\cite{Yamakami20}, lambda-calculi~\cite{DALLAGO2010377} and more recently programming languages~\cite{HPS23}. This last work also made the first explicit connection between a characterization of quantum polynomial time and polynomial-size quantum circuit families.

\paragraph*{Quantum recursive branching.}
The complexity of quantum branching in the context of recursion has been observed in other work, for instance in~\cite{TOWER} where recursive quantum branching is explicitly avoided in program design so as to prevent an exponential slow-down in the compilation, a problem the authors coin as ``branch sequentialization.''

In the study of tackling branch sequentialization, a recent work~\cite{ZY25} develops a number of techniques that aim at parallelizing quantum recursive procedure calls so as to reduce circuit depth. This is done automatically by the compiler (that is to say, the programmer does not need to explicitly set up this parallelization in the program) and the resulting circuit depth of quantum control is, up to a linear additive factor, the maximum depth of each branch. 

\paragraph*{Static analysis.} The fact that the restrictions on programs are of a syntactic nature allows for the complexity to be verified statically, and in turn provides tools for the programmer to more easily and flexibly reason about their program's complexity. Implicit characterizations also allow for directly considering the balance between the strength of results (how much one can say about a program that follows certain syntactical restrictions) and the expressivity of the characterization (how easy it is to write programs that follow those restrictions).

In classical programming languages, certifying programs for their resource usage has inspired the development of different software tools~\cite{complexityparser,mwp2022}, with techniques directly inspired by implicit computational complexity.

\section{First-order quantum programs}

In this section, we recall the first-order quantum (\foq{}) programming language introduced in~\cite{HPS23} and expanded in~\cite{FHPS25}. We describe its syntax and operational semantics.

\begin{table*}[t]
{\ttfamily
\begin{tabular}{rrll}
     \textnormal{Proc. decl.} & D\ :=&  $\epsilon \mid$ \tbt{decl }\pn{proc}[x,...,x](a,\dots,a)\{S\}, D  & \\
     \textnormal{Statements}  & S\ := & \tbt{skip}; & \textnormal{(no-op instruction)}\\
     & & $\mid$\ l[i]\ *=\ U(j); &\textnormal{(gate application)}\\
     & & $\mid$\ CNOT(q,q); $\mid$ TOF(q,q,q); $\mid$ SWAP(q,q);\\
     & & $\mid$\ S\ S & \textnormal{(composition)}\\
     & & $\mid$\ \tbt{if} b\tbt{ then }\{S\}\tbt{ else }\{S\} & \textnormal{(classical if)}\\
     & & $\mid$\ \tbt{qcase }(l[i])\tbt{ of }\{0 -> S,\,1 -> S\} & \textnormal{(quantum if)}\\
     & & $\mid$\ \tbt{call }\pn{proc}[i,...,i](l,\dots,l); & \textnormal{(procedure call)}\\
     \textnormal{Unitaries}  & U\ :=& H $\mid$ NOT $\mid$  Rot[f] $\mid$ Ph[f]\textnormal{,$\quad$}f\textnormal{$\ \in \mathbb{Z}\to\mathbb{R}$} &\\
     \textnormal{Integers}& i, j\ :=& x $\mid$ $n\in\mathbb{Z}$ $\mid$ |l| $\mid$ i + $n$ $\mid$ i - $n$ $\mid$ i//2\\\textnormal{Booleans}& b\ :=& \tbt{true} $\mid$ \tbt{false} $\mid$ compare(i,j)\\
     && $\mid$ \tbt{!}b $\mid$ b \tbt{\&\&} b $\mid$ b \tbt{||} b \\
     \textnormal{Qubit lists} & l\ :=& a $\mid$ q $\mid$  l - [i,\dots,i] $\mid$ (l)\^{}- $\mid$ (l)\^{}+
\end{tabular}
}
\caption{Syntax of \foq{}.}
\label{table:syntax}
\end{table*}

\subsection{Syntax}

The syntax for \foq{} programs is given in Table~\ref{table:syntax}. A \foq{} program consists of three parts:
\[\texttt{D ::\ \tbt{define }q$_1\ \dots\ $\texttt{q}$_k$ ::\ \ST}\]

The list of procedure declarations \texttt{D} contains the definition of all the (possibly recursive) procedures that can be called in the program. These procedures are defined over qubit list variables of the form \texttt{a}, \texttt{b}, \texttt{c},  \texttt{a0}, \texttt{a1}, \dots \texttt{a100} and so on. The program then defines the names of the different qubit lists given as input to the program, which constitute the \emph{variables of the program} \texttt{P}, denoted $Var(\PR)=\{\texttt{q}_1,\dots,\texttt{q}_k\}$. Finally, the program statement \ST{} contains the program statement to be executed.

The most basic statements are the no-op instruction \tbt{skip}, and the unitary gate application \texttt{q[i]*= U[f](j)}, which applies a single-qubit gate $U(f,j)$ to the $i$-th element of the qubit list \texttt{q}. The Hadamard and NOT gates are denoted respectively as \texttt{H} and \texttt{NOT}, and correspond to the following gates:
\[\text{H}\triangleq \frac{1}{\sqrt{2}}\begin{pmatrix} 1 & 1\\ 1 & -1\end{pmatrix},\quad \text{NOT}\triangleq \begin{pmatrix} 0 & 1\\ 1 & 0\end{pmatrix}.\]
Other gates such as the phase-shift and rotation gates take in a function and a classical input parameter that determines the angle of the shift or rotation. For instance, the gate \texttt{Ph[f](i)} corresponds to the phase-shift gate
\[ \text{P}(f,i) \triangleq \begin{pmatrix}
1 & 0\\
0 & e^{\imath \cdot f(i)}
\end{pmatrix}
\]
and the rotation gate \texttt{Rot[f](i)} is given by the $y$-rotation:
\[
\text{R}_y(f,i) \triangleq
\begin{pmatrix}
\cos(f(i)) & -\sin(f(i))\\
\sin(f(i)) & \cos(f(i))
\end{pmatrix}.
\]

\foq{} also allows for the use of classical and quantum control: given a classical boolean condition \texttt{b} evaluating to $b\in\{\tbt{true},\,\tbt{false}\}$, the statement
\[\texttt{\tbt{if }(b)\tbt{ then }\{S$_\tbt{true}$\}\tbt{ else }\{S$_\tbt{false}$\}}\]
executes the set of instructions $\texttt{S}_b$. Notice that the value of $b$ can only depend on classical information that can be known at compile time (since \foq{} does not include qubit measurements). If $\texttt{S}_\tbt{false}$ is the no-op instruction, then the \tbt{else} branch can be omitted.

In the case of quantum control, the guard is given by a qubit address $\texttt{l[i]}$ and the two branches correspond to coherent control according to the state of the qubit. The instruction
\[\texttt{\tbt{qcase }(l[i])\tbt{ of }\{0 -> S$_0$,\,1 -> S$_1$\} }\]
executes statements \texttt{S$_0$} and \texttt{S$_1$} on the subspaces where the qubit corresponding to address \texttt{l[i]} is in state $0$ or $1$, respectively.

Multiple nested \tbt{qcase} statements can be used to perform quantum control over many qubits at a time. As syntactic sugar, we introduce the following 2-qubit \tbt{qcase} statement:
\vspace{3mm}

\begin{PFOQprogram}
\tbt{qcase }(l$_1$[i],l$_2$[j])\tbt{ of }\{ 00 -> S$_{00}$,\\
\hphantom{\tbt{qcase }(l$_1$[i],l$_2$[j])\tbt{ of }\{} 01 -> S$_{01}$,\\
\hphantom{\tbt{qcase }(l$_1$[i],l$_2$[j])\tbt{ of }\{} 10 -> S$_{10}$,\\
\hphantom{\tbt{qcase }(l$_1$[i],l$_2$[j])\tbt{ of }\{} 11  ->  S$_{11}$\ \}\\
\end{PFOQprogram}
\vspace{3mm}

\noindent
The \texttt{CNOT} and \texttt{TOF} statements corresponding to the $CNOT$ and Toffoli gates can then be straightforwardly defined from the 1- and 2-qubit \tbt{qcase}. 

We give a number of illustrative examples of \foq{} programs in Section~\ref{s:examples}.

\subsection{Program time complexity}
\label{ss:semantics}

The main motivation of the \pfoqcompiler{} is to generate families of circuits that faithfully correspond (both in semantics and complexity) to the \foq{} program in question, thereby avoiding the problem of branch sequentialization.

For this reason, we must use a notion of \emph{time complexity} of a program which captures its asymptotic behavior. Since the only non-constant-time instruction is the procedure call, a straightforward way of defining the asymptotic time complexity of a \foq{} program is the maximum number of procedure calls performed in depth, where we consider the complexity of classical or quantum branching to be the \emph{maximum} time complexity of the branches.

This notion of time complexity is defined precisely in~\cite{HPS23,FHPS25} using operational semantics. For a program $\PR$ and input sizes $\mathcal{N}\triangleq n_1,\, \dots, n_k$, we denote the time complexity of the program as the function \(\text{Time}_\PR(\mathcal{N})\).

The time complexity of a program can usually be gleaned straightforwardly from the program structure. For example, in \texttt{\color{gray}PAIRS} (Figure~\ref{fig:pairs}), each instance of \pn{pairs} performs one recursive call (maximum over the classical and quantum branches). Therefore, one can easily conclude that $\text{Time}_{\texttt{\color{gray}PAIRS}}(n)=\mathcal{O}(n)$.
 
\paragraph{Semantics.} We describe semantics of \foq{} statements by example in Section~\ref{s:examples}. We give here a precise meaning to \foq{} expressions over integers and qubit lists, as they differ the most from standard programming languages. A full \foq{} operational semantics is described in~\cite{FHPS25}.

Given a list \texttt{l} of qubit addresses, say $[x_1,\dots,x_m]$, then \texttt{(l)\^{}-} denotes the first half of \texttt{l}, defined as $[x_1,\dots,x_{\lceil \sfrac{m}{2}\rceil}]$, and \texttt{(l)\^{}+} denotes the remaining half, i.e. $[x_{\lceil \sfrac{m}{2}\rceil+1},\dots,x_m]$. If the list contains a single qubit address (that is to say, $m=1$), then both halves are defined to be the empty set. The integer value \texttt{i//2} is defined as $\lceil \texttt{i}/2\rceil$.

Two important properties of \foq{} semantics that play a role in the complexity analysis are the following:
\begin{enumerate}
\item if a procedure takes in \emph{at least} one empty qubit list as input, then the procedure halts without performing any operation;
\item all operations on a qubit list either strictly decrease its size or result in an error.
\end{enumerate}

Behavior 1 implies that a procedure defined as follows:
\vspace{3mm}

\begin{PFOQprogram}
\tbt{decl }\pn{had}(a,b)\{\\
\quad a[0] *= H;\\
\quad \tbt{call }\pn{had}(a-[0],b); \}
\end{PFOQprogram}
\vspace{3mm}

\noindent is well-defined for any input size as either the input list \texttt{a} has a first element \texttt{a[0]} or it is empty, in which case the procedure safely halts.

Importantly, \pn{had} also halts if the input list \texttt{b} is empty, even if it does not make use of its second argument. This ensures that one may check that \pn{had} reduces the size of at least one of its input qubit lists (in this case, the first) to demonstrate that \pn{had} always halts.

Behavior 2 has a similar motivation, as the objective is to ensure that an expression of the form
\[\texttt{l-[i,...,i]}\quad\text{or}\quad\texttt{(l)\^{}-}\quad\text{or}\quad\texttt{(l)\^{}+} \]
always results in a new list whose size is strictly smaller than that of \texttt{l}, whenever the program does not produce an error. For this reason, the first or second halves of a list containing a single element are both the empty list.

\section{\pfoqcompiler{} by example}
\label{s:examples}

We now demonstrate the \pfoqcompiler{} through some illustrative examples.

\paragraph{Quantum control flow.}

Consider the following example that describes the quantum circuit for creating a Bell state
\[\frac{1}{\sqrt{2}} \big(\ket{00} + \ket{11}\big) \]
from a $\ket{00}$ input state. It consists of applying a Hadamard gate to the first qubit, followed by a CNOT gate controlled on the first qubit and targeting the second one.
\vspace{3mm}

\noindent
\begin{minipage}{0.45\columnwidth}
\scalebox{0.9}{
\begin{PFOQprogram}
::\\
\tbt{define }q;\\
::\\
q[0] *= H;\\
\tbt{qcase }(q[0])\tbt{ of }\{\\
\quad 0 -> \tbt{skip};,\\
\quad 1 -> q[1] *= NOT; \}
\end{PFOQprogram}}
\end{minipage}
\begin{minipage}{0.5\columnwidth}
\begin{center}
\scalebox{1}{
\includegraphics[scale=0.45,trim={20mm 10mm 10mm 10mm},clip]{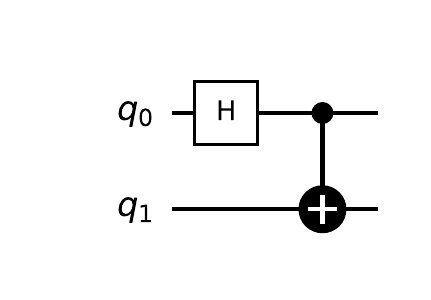}}
\end{center}
\end{minipage}
\vspace{3mm}

The CNOT gate is implemented in the style of a general controlled gate. Since this is a very common instruction, we include it as  a basic operation in the syntax. Therefore, the following program results in precisely the same circuit.
\vspace{3mm}

\noindent
\begin{minipage}{\columnwidth}
\centering
\scalebox{0.9}{
\begin{PFOQprogram}
::\\
\tbt{define }q;\\
::\\
q[0] *= H;\\
CNOT(q[0],q[1]);
\end{PFOQprogram}}
\vspace{3mm}
\end{minipage}

The Toffoli gate (i.e. the 2-controlled $NOT$) can also be used directly using the $\texttt{TOF}$ command.

\paragraph{Iteration.}

The above examples can only produce a two-qubit Bell state from the input. In order to create a larger $n$-qubit GHZ state, we must iterate over the input qubit list. In \foq{}, this is done using recursion, as in the following example.
\vspace{3mm}

\noindent
\begin{minipage}{0.5\columnwidth}
\scalebox{0.8}{
\begin{PFOQprogram}
{\color{gray}GHZ}\\
\\
\tbt{decl }\pn{cnots}(a)\{\\
\quad \tbt{if }(|a|>1)\tbt{ then }\{\\
\qquad CNOT(a[0],a[1]);\\
\qquad \tbt{call }\pn{cnots}(a-[0]);\}\,\}\\
::\\
\tbt{define }q;\\
::\\
q[0] *= H;\\
\tbt{call }\pn{cnots}(q);
\end{PFOQprogram}}
\end{minipage}
\begin{minipage}{0.45\columnwidth}
\begin{center}
\includegraphics[scale=0.3,trim={26mm 18mm 16mm 20mm},clip]{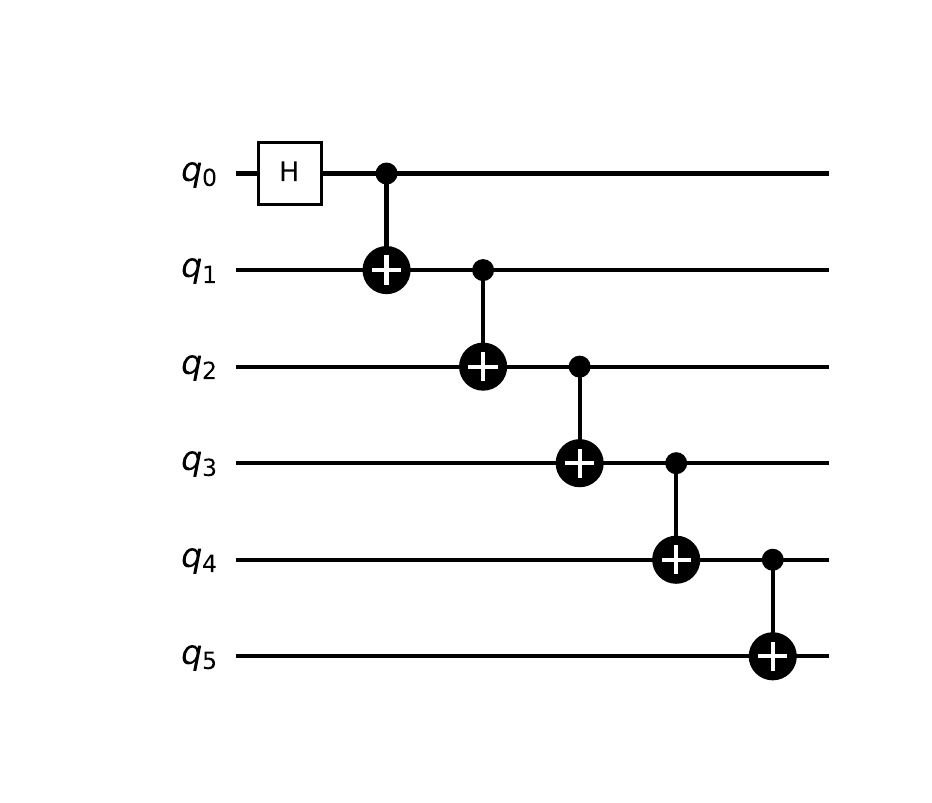}
\end{center}
\end{minipage}
\vspace{3mm}

An implementation of GHZ creation in parallel can also be described in \foq{}, as in the following example, where the CNOT gates propagate in parallel in a circuit of logarithmic depth~\cite{GHZ_efficient}.

\vspace{3mm}

\noindent
\begin{minipage}{0.79\columnwidth}
\scalebox{0.8}{
\begin{PFOQprogram}
{\color{gray}GHZ\_PAR}\\
\\
\tbt{decl }\pn{cnots\_par}(a)\{\\
\quad \tbt{if }(|a|>2)\tbt{ then }\{\\
\qquad  CNOT(a[0],a[|(a)\^{}-|]);\\
\qquad  \tbt{call }\pn{cnots\_par}((a)\^{}-);\\
\qquad  \tbt{call }\pn{cnots\_par}((a)\^{}+); \}\\
\quad \tbt{else }\{\\
\quad \tbt{if }(|a|==2) \tbt{then }\{\\
\qquad CNOT(a[0],a[1]);\,\}\ \}\ \}\\
::\\
\tbt{define }q;\\
::\\
q[0] *= H;\\
\tbt{call }\pn{cnots\_par}(q);
\end{PFOQprogram}}
\end{minipage}
\begin{minipage}{0.18\columnwidth}
\begin{center}
 \includegraphics[scale=0.35,trim={390mm 20mm 340mm 25mm},clip]{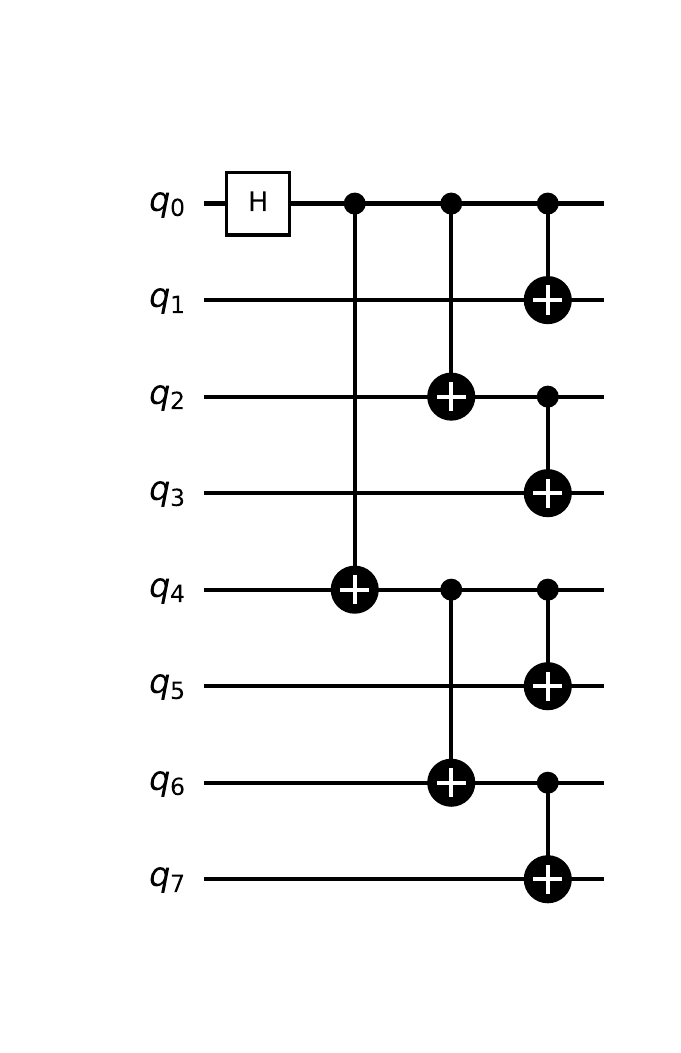}
\end{center}
\end{minipage}

\paragraph{Integer variables and parametrized gates.}
\label{ss:parametrized-gates}

Two types of variables can be passed to procedures: classical integer variables or lists of quantum addresses. We show how integer variables can be used in the following toy example.
\vspace{3mm}

\noindent
\scalebox{0.78}{
\begin{PFOQprogram}
{\color{gray}SHIFTS}\\
\\
\tbt{decl }\pn{shifts}[x](a,b)\{\\
\quad a[0]\asg H;\\
\quad \tbt{qcase }(a[0])\tbt{ of }\{\\
\qquad 0 -> \tbt{skip};\\
\qquad 1 -> b[0]\asg Ph[lambda x:pi/(2**x)](x);\}\\
\quad \tbt{call }\pn{shifts}[x+1](a-[0],b);
\}\\
::\\
\tbt{define }q r;\\
::\\
\tbt{call }\pn{shifts}[0](q,r);
\end{PFOQprogram}}
\vspace{0mm}
\begin{center}
 \includegraphics[scale=0.33,trim={50mm 20mm 10mm 20mm},clip]{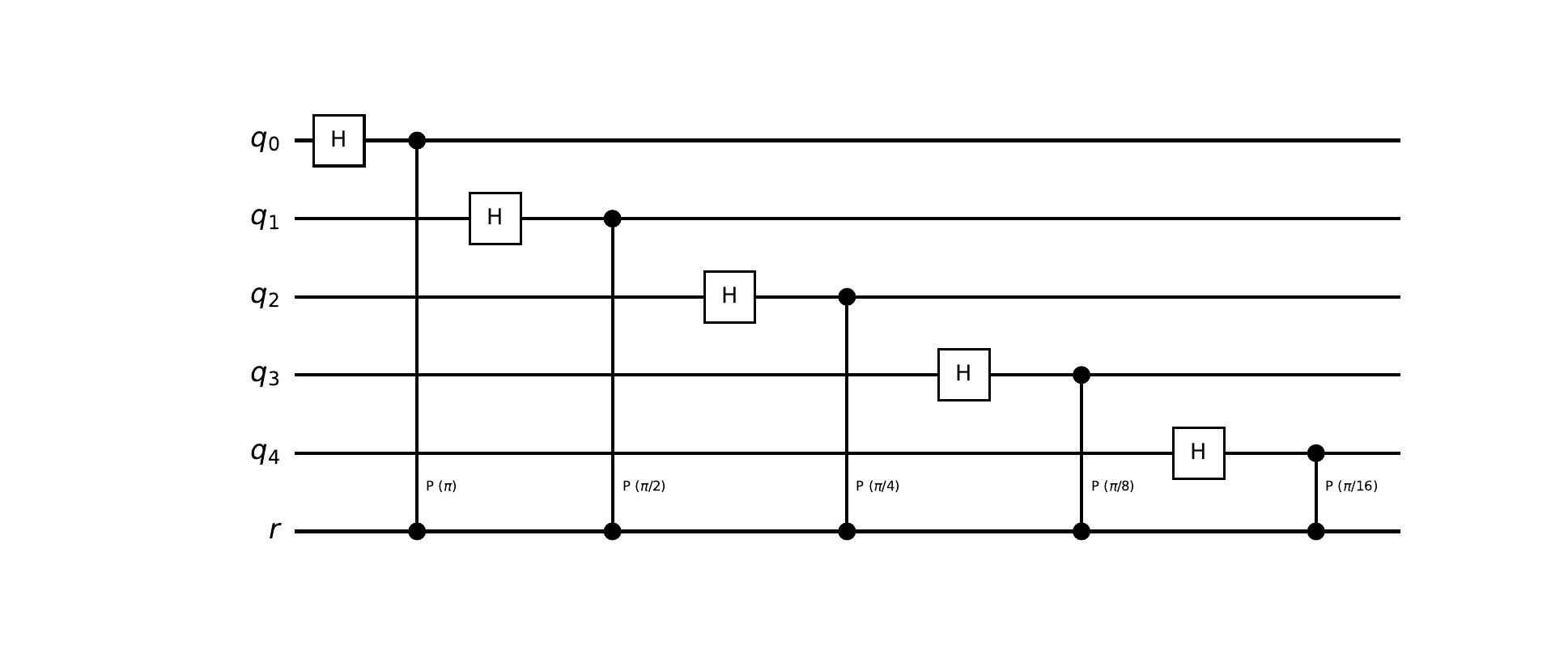}
\end{center}

\begin{figure*}[t]
\begin{center}
 \includegraphics[width=\textwidth,trim={115mm 15mm 60mm 10mm},clip]{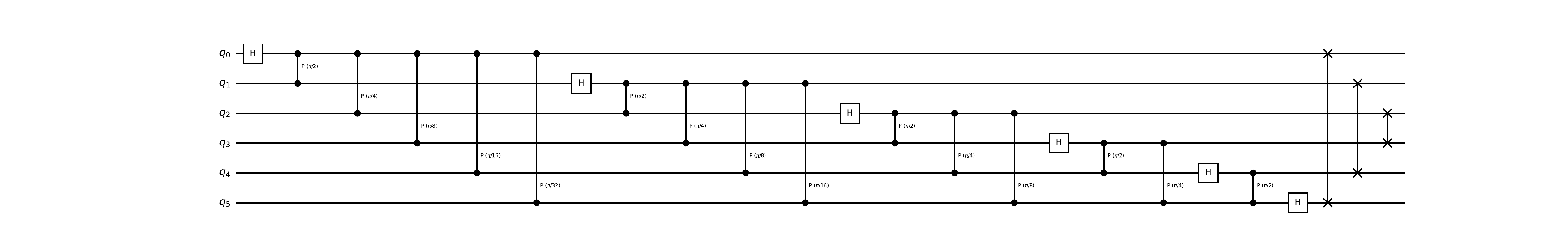}
\caption{Circuit for the quantum Fourier transform on input size 6.}
\label{fig:qft}
\end{center}
\end{figure*}

\paragraph*{Quantum Fourier transform.} We give a \foq{} program for the quantum Fourier transform~\cite{QFT}, where procedure \pn{rot} implements the sequence of controlled phase-shift gates (Figure~\ref{fig:qft}) using a parametrized phase-shift gate \texttt{Ph}. The first input to the gate is the function $\lambda x.\sfrac{\pi}{2^x}$, a second argument $\texttt{|a|-1}$ defines the integer that is passed on as input to the function and the result is the angle of the phase-shift.
\vspace{3mm}

\noindent
\scalebox{0.78}{
\begin{PFOQprogram}
{\color{gray}QFT}\\
\\
\tbt{decl }\pn{qft}(a)\{\\
\quad a[0] *= H;\\
\quad \tbt{call} \pn{rot}(a);\\
\quad \tbt{call} \pn{qft}(a-[0]); \}\\
\\
\tbt{decl }\pn{inv}(a)\{\\
\quad \tbt{if }(|a|>1)\tbt{ then }\{\\
\qquad	SWAP(a[0],a[-1]);\\
\qquad	\tbt{call }\pn{inv}(a-[0,-1]);\}\}\\
\\
\tbt{decl }\pn{rot}(a)\{\\
\quad \tbt{if} (|a|>1) \tbt{then }\{\\
\qquad \tbt{qcase} (a[1])\tbt{ of }\{\\
\qquad \quad 0 -> \tbt{skip};,\\
\qquad \quad 1 -> a[0] *= Ph[lambda x:pi/(2**x)](|a|-1);\}\\
\qquad \tbt{call }\pn{rot}(a-[1]);\} \}\\
::\\
\tbt{define }q;\\
::\\
\tbt{call }\pn{qft}(q);
\tbt{call }\pn{inv}(q);
\end{PFOQprogram}}

\paragraph{Quantum ripple-carry adder.}

The following program implements the quantum ripple-carry adder of registers $\texttt{q}$ and $\texttt{p}$ into register $\texttt{r}$~\cite{Vedral_1996}.

\vspace{3mm}

\noindent
\scalebox{0.85}{
\begin{PFOQprogram}
\tbt{decl }\pn{qrca}(a,b,c)\{\\
\quad \tbt{if }(|a|>1 \tbt{\&\&} |b|>1 \tbt{\&\&} |c|>2)\tbt{ then }\{\\
\qquad TOF(a[-1],b[-1],c[-2]);\\
\qquad	CNOT(a[-1],b[-1]);\\
\qquad	TOF(b[-1],c[-1],c[-2]);\\
\qquad	CNOT(b[-1],c[-1]);\\
\qquad	CNOT(a[-1],b[-1]);\\
\qquad \tbt{call }\pn{qrca}(a-[-1],b-[-1],c-[-1]);\}\ \}\\
::\\
\tbt{define }q p r;\\
::\\
\tbt{call }\pn{qrca}(q,p,r);
\end{PFOQprogram}}
\vspace{0mm}
\begin{center}
 \includegraphics[scale=0.33,trim={40mm 20mm 28mm 27mm},clip]{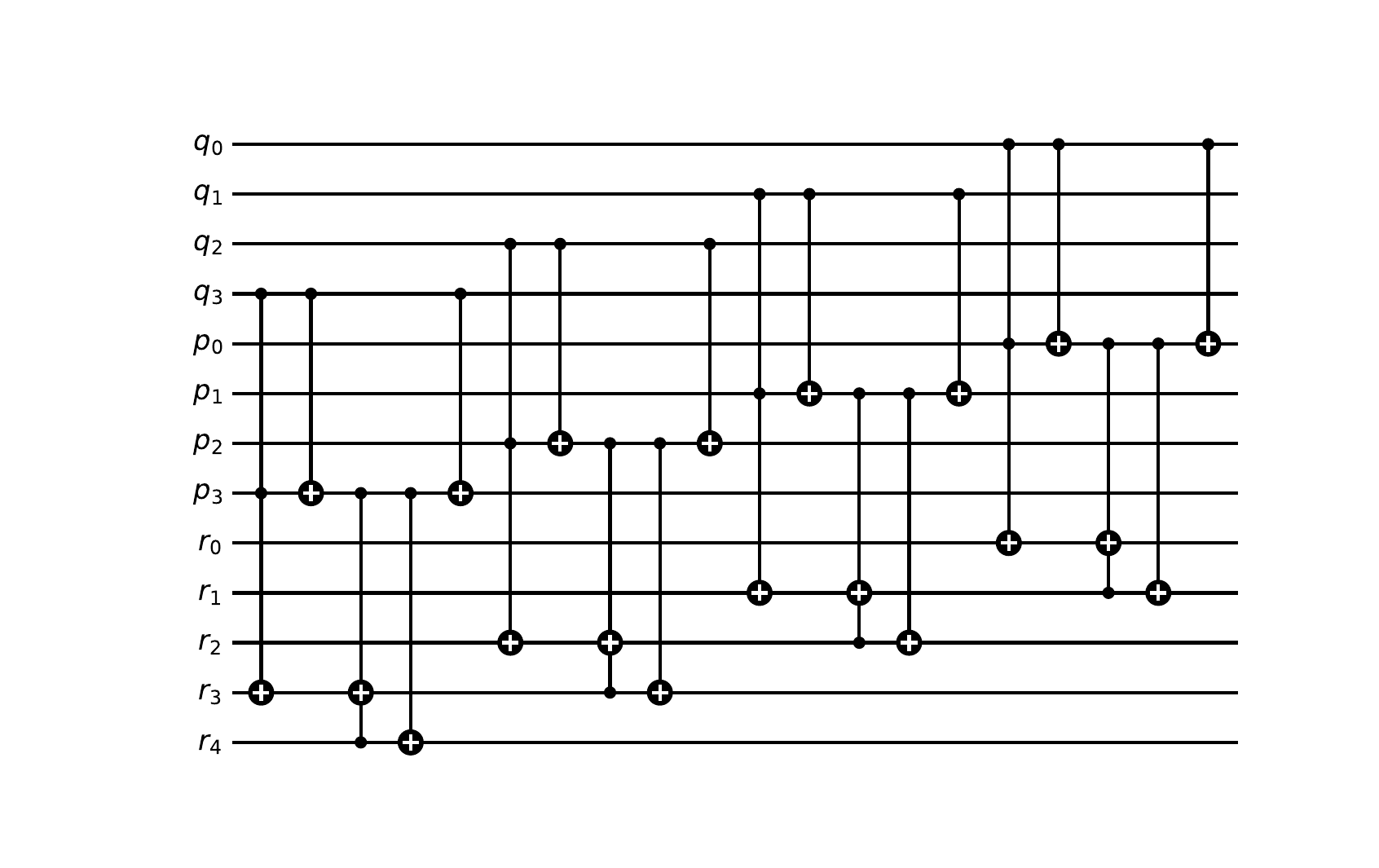}
\end{center}

\paragraph{Merging procedure calls.}

We revisit the \pn{pairs} example of the introduction to show the circuit obtained by the merging strategy.
\vspace{3mm}

\noindent
\begin{minipage}{0.45\columnwidth}
\scalebox{0.85}{
\begin{PFOQprogram}
{\color{gray}PAIRS}\\
\\
\tbt{decl }\pn{pairs}(a)\{\\
\quad \tbt{if }(|a|>1)\tbt{ then }\{ {\color{gray} \# recursive case}\\
\qquad  \tbt{qcase }(a[0],a[1])\tbt{ of }\{ \\
\qquad\quad  00 -> \tbt{call }\pn{pairs}(a-[0,1]);,\\
 \qquad\quad 01 -> \tbt{skip};,\\
 \qquad\quad 10 -> \tbt{skip};,\\
\qquad\quad  11  -> \tbt{call }\pn{pairs}(a-[0,1]);\ \}\\
\quad  \} \tbt{else} \{ a[0] *= NOT; \}\ \} {\color{gray} \# base case}\\
::\\
\tbt{define }q;\\
::\\
\tbt{call }\pn{pairs}(q);
\end{PFOQprogram}}
\end{minipage}
\begin{minipage}{0.55\columnwidth}
\begin{center}
\vspace{3mm}
 \includegraphics[scale = 0.23,trim={70mm 0 50mm 10mm},clip]{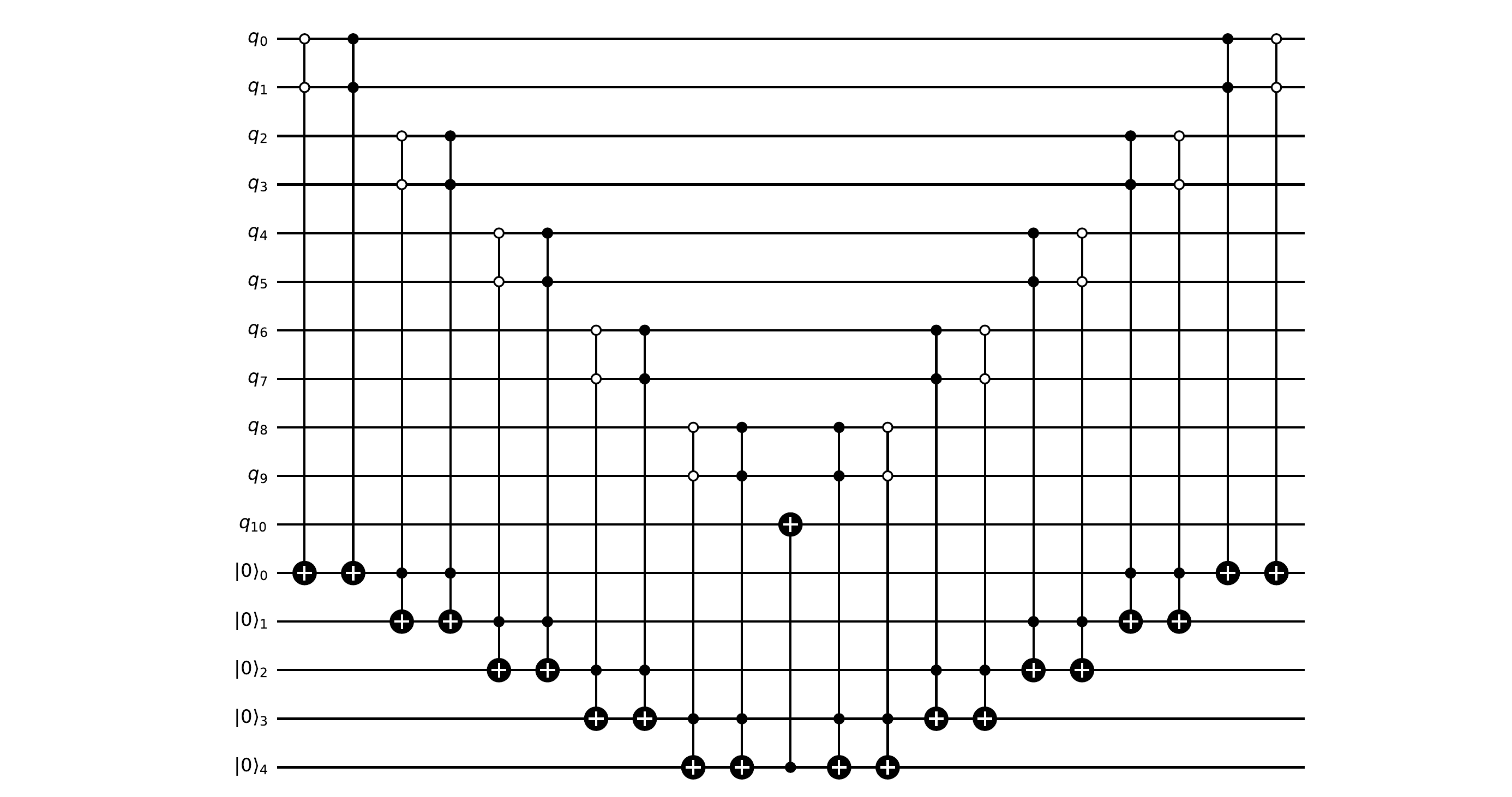}
\end{center}
\end{minipage}
\vspace{3mm}

\paragraph*{Binary search.}
We now consider a binary-search program. Let \texttt{q} encode a sorted string $x\in\{0,1,2\}^n$ such that $\hat{x}=\hat{x}_1\hat{x}_2\dots \hat{x}_n$ is $x$ encoded into binary such that \(\hat{0} = 00\), \(\hat{1} = 01\), and \(\hat{2} = 10\). Then, the following program takes in \texttt{q} and an output register \texttt{r} such that the first qubit of \texttt{r} is flipped if $1\in x$.

For an input size \texttt{|q|} of 6, five total calls to \pn{search} are performed: the first one, on the entire input, i.e. \texttt{q[0,...,5]}, then two recursive calls are performed on the first and second halves of \texttt{q}, excluding the middle. They correspond, respectively, to \texttt{q[0,1]} and \texttt{q[4,5]}. Given that these two instances of \pn{search} are compatible, only one is compiled (in this case, the one with input \texttt{q[4,5]}) and the other is merged. This instance of \pn{search} on input size 2 then results in two more calls to \pn{search} on empty inputs, which terminate as the identity.

\begin{minipage}{0.45\columnwidth}
\scalebox{0.85}{
\begin{PFOQprogram}
{\color{gray}SEARCH}\\
\\
\tbt{decl }\pn{search}(a,b)\{\\
\quad \tbt{if }(|a|>1)\tbt{ then }\{\\
\qquad   \tbt{qcase }(a[|a|/2-1], a[|a|/2])\tbt{ of }\{\\
\qquad\quad   00 -> \tbt{call }\pn{search}((a)\^{}+ -[0],b);,\\
\qquad\quad 01 -> b[0] *= NOT;,\\
\qquad\quad  10 -> \tbt{call }\pn{search}((a)\^{}- -[-1],b);,\\
\qquad\quad  11 -> \tbt{skip}; \} \} \} \\
::\\
\tbt{define }q r;\\
::\\
\tbt{call }\pn{search}(q,r);
\end{PFOQprogram}}
\end{minipage}
\vspace{3mm}

\begin{minipage}{0.55\columnwidth}
\begin{center}
 \includegraphics[scale=0.3,trim={52mm 25mm 35mm 30mm},clip]{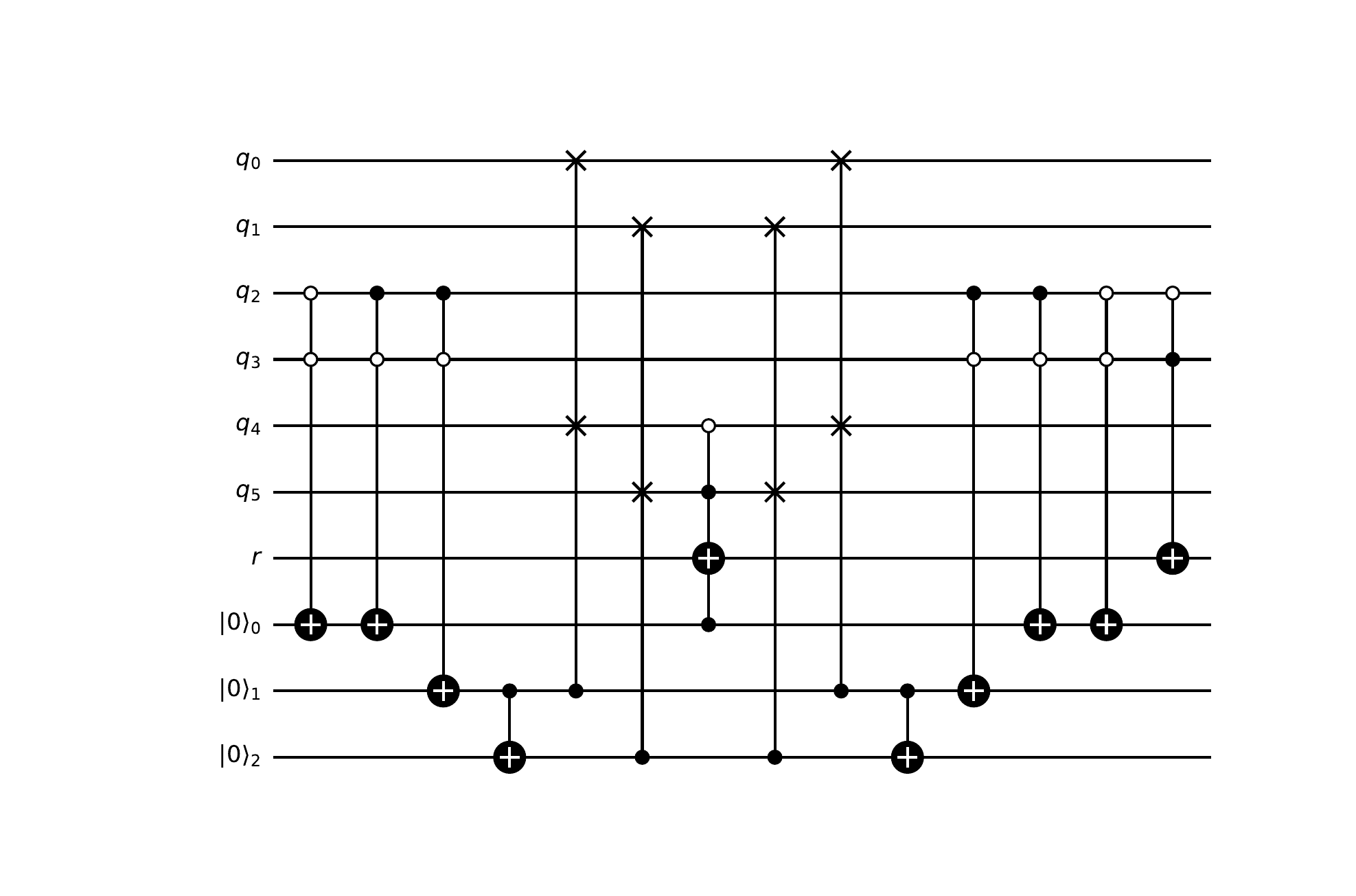}
\end{center}
\end{minipage}
\vspace{3mm}

\section{Syntactical restrictions and static verification}\label{sec:restrictions}

The \texttt{pfoqcompiler} reduces the asymptotic complexity of quantum branching by merging procedure calls whenever possible, much in the same way as described in the introduction for the \pn{pairs} procedure. This circuit optimization is only valid conditionally on the structure of the program, as it is precisely this structure that is exploited to perform simplifications.

In this section, we introduce the restrictions used for static verification of programs in~\cite{HPS23,FHPS25,HPS25} to ensure complexity bounds on circuit compilation. We denote these restrictions as \emph{strong} and we introduce \emph{weak} versions that allow for more expressivity.

We denote by $\mathcal{I}$ a list of integers, and by $\mathcal{L}$ a list of qubit address lists.

\paragraph*{Well-structured programs.} 

Given our intent in simplifying procedure calls, we impose syntactic restrictions which ensure the conditions for merging procedures during compilation.

Two procedure calls can be merged if they encode the same subcircuit in the program. Such procedure calls are called \emph{compatible} and are determined by the procedure name, and their classical and quantum inputs.

\begin{definition}[Procedure compatibility]
Two procedure calls \(\tbt{call }\pn{proc}\textnormal{\texttt{[$\mathcal{I}$]($\mathcal{L}$)}}\) and
\(\tbt{call }\pn{proc'}\textnormal{\texttt{[$\mathcal{I}'$]($\mathcal{L}'$)}}\)
are said to be \emph{compatible} if
\begin{align*}
\pn{proc}=\pn{proc'},\quad\mathcal{I}'=\mathcal{I},\quad\textnormal{and}\quad |\mathcal{L}|=|\mathcal{L}'|,
\end{align*}
where the size comparison between $\mathcal{L}$ and $\mathcal{L}'$ is done pairwise.
\end{definition}

Put simply, compatible procedures are those for which the procedure names and integer inputs match precisely, and the quantum inputs match in length. This fixes the classical control flow of the procedures.

If two (or more) compatible procedure calls occur in different quantum branches of computation, the anchoring and merging compilation technique will simplify these procedure calls into a single one, which prevents the recursive blow-up of branch sequentialization~\cite{TOWER}.

Given a list of qubit variables \texttt{l}, we define $var(\texttt{l})$ as the variable from which \texttt{l} is derived. For instance, $var(\texttt{q-[|r|]})=\texttt{q}$.

To ensure that compatibility between procedure calls can be statically checked, we require that programs are \emph{well-structured}, by the following definition. 

\begin{definition}[Well-structured programs]\label{def:well-structured} A \foq{} program $\PR$ is defined as \emph{well-structured} if it satisfies the following conditions:
\begin{enumerate}
\item Procedures in $\PR$ make no use of global variables.
\item  All procedure calls occurring in $\PR$ have disjoint input lists of qubits, i.e.
$\forall\, \tbt{call }\proc\textnormal{\texttt{[$\mathcal{I}$](l$_1$,...,l$_k$)}}\in\PR$ and $\forall i\neq j$, $var(\textnormal{\texttt{l}}_i) \neq var(\textnormal{\texttt{l}}_j)$.
\end{enumerate}
\end{definition}

Both these conditions ensure that compatible procedure calls are described by the same unitary transformation, up to a swap of input qubits.

To motivate these restrictions, we show a counterexample for each. One important note is that these restrictions do not serve to guarantee error-free programs, but rather to avoid errors that could be ignored by the anchoring-and-merging technique.

\paragraph{Restriction on global variables.}\label{app:global-variables}

The first restriction we consider is the fact that procedures cannot make use of global qubit list variables. Consider the following procedure \proc, where $\texttt{q}$ denotes a global list of qubits.

\begin{center}
\scalebox{1.}{
\begin{PFOQprogram}
\tbt{decl }\pn{proc}(a)\{ CNOT(a[0],q[0]);\,\}
\end{PFOQprogram}}
\end{center}

If \proc{} ever gets called on qubit list \texttt{q}, the \texttt{CNOT} operation is ill-defined and the program results in an error. However, if this procedure appears in a \foq{} program with more than one qubit list, say \texttt{q} and \texttt{r}, then procedure \proc{} can be called on qubit list \texttt{r} without issue.

Now, if we consider the specific case where \texttt{q} and \texttt{r} have the same length, this raises a problem for the compilation strategy. The two calls should be compatible, as they pertain to the same procedure, there are no classical inputs to compare, and the input lists have the same size. However these procedures are not semantically equivalent, as one executes correctly and the other terminates in error.

Notice that it is not required to have different qubit lists to find such a counterexample, as for instance $\proc$ executes correctly on input \texttt{q-[0]} and fails on input \texttt{q-[1]} which has the same size.

\paragraph{Disjoint input qubit lists.} 

We now turn to the second restriction, namely that which enforces that procedure calls must use disjoint input qubit lists. This is statically verified by checking that input qubits have different qubit list variables, for instance,
\[\tbt{call }\proc\texttt{(q, r - [|q|]);}\]
is allowed, whereas
\[\tbt{call }\proc\texttt{(q, q - [1]);}\]
is not.

Even in cases where the program executes successfully, this breaks the validity of the merging technique. Consider, for instance, a procedure \proc{} defined as follows:

\vspace{2mm}

\begin{PFOQprogram}
\tbt{decl }\pn{proc}(a,b)\{\\
\quad a[0] *= H;\\
\quad b[0] *= NOT;\}
\end{PFOQprogram}\\
 
The procedure \proc{} is valid for whichever pair of qubit lists is given as input. However, given qubit lists \texttt{q} and \texttt{r}, even though both
 \[\tbt{call }\proc\texttt{(q,q);}\qquad
\begin{quantikz}[row sep = {8mm, between origins}]
\lstick{\texttt{q}} & \gate{H} & \gate{X} &\\
\lstick{\texttt{r}} &  &  &
\end{quantikz} 
 \]
 and 
 \[
\tbt{call }\proc\texttt{(q,r);}\qquad\qquad
\begin{quantikz}[row sep = {8mm, between origins}]
\lstick{\texttt{q}} & \gate{H} & \\
\lstick{\texttt{r}} & \gate{X} &  
\end{quantikz}  
 \]
are safe in the sense that they correspond to valid circuits, they \emph{do not} correspond to the same circuit up to a swapping of qubit addresses, and therefore in the case where the sizes of \texttt{q} and \texttt{r} are equal they cannot be merged.

\paragraph*{Mutual recursion.}
In order to restrict recursive procedure calls, we give a precise definition of mutual recursion in \foq{} programs, by defining the relations \[\geq_\PR,\, \succeq_\PR, \sim_\PR, \succ_\PR\ \in\ \text{Procedures}\times \text{Procedures}.\] Let \pn{proc}, \pn{proc'} be two procedures defined in a \foq{} program \PR. We write $\pn{proc}\geq_\PR\pn{proc'}$ if $\tbt{call }\pn{proc'}\in\ST^\pn{proc}$. Let $\succeq_\PR$ be the preorder defined as the reflexive and transitive closure of $\geq_\PR$, and let $\sim_\PR$ be the equivalence relation defined by $\pn{proc}\sim_\PR\pn{proc'}$ if both $\pn{proc}\succeq_\PR\pn{proc'}$ and $\pn{proc'}\succeq_\PR\pn{proc}$ hold. Finally, let $\pn{proc}\succ_\PR \pn{proc'}$ be defined as the strict case where $\pn{proc}\succeq_\PR \pn{proc'}$ but $\pn{proc}\not\sim_\PR \pn{proc'}$. The subscript \PR{} can be dropped if it is clear from context.

For instance, in the quantum Fourier program \texttt{\color{gray}QFT} of Section~\ref{s:examples}, we have that
\[ \pn{qft}\sim \pn{qft}\succ \pn{rot} \sim \pn{rot}\quad\text{and}\quad \pn{inv} \sim \pn{inv}.\]

Note that by reflexivity of $\succeq$ we have that $\pn{proc}\sim\pn{proc}$ even if \pn{proc} is not recursive.

\paragraph*{Termination.} By language design, programs necessarily terminate once one (or more) of the input qubit lists is empty. Therefore, termination can be guaranteed by checking that (mutually) recursive calls always reduce the size of (at least) one of their input qubit lists. This property is called \emph{well-foundedness}.

\begin{definition}[Strong well-foundedness, adapted from~{\cite[Definition 1]{HPS23}}]
A \foq{} program \PR{} is said to be \emph{strongly well-founded} (written $\PR\in\wf{}$) if, for every pair of procedures $\proc,\, \pn{proc'} \in\PR$ such that $\proc\sim_\PR \pn{proc'}$, any procedure call to $\pn{proc'}$ appearing in $\textnormal{\ST}^\proc$ reduces the number of elements in some input qubit list by at least 1.
\end{definition}

Notice that $\PR\in\wf{}$ can be statically checked by inspecting $\PR$ without running the program, given \foq{}'s restricted syntax. We now introduce a \emph{weaker} form of well-foundedness.

\begin{definition}[Weak well-foundedness] \label{def:weak-wf} A \foq{} program \PR{} is said to be \emph{weakly well-founded} (written $\PR\in\wf{}^\star$) if there is no sequence of procedure calls
\[\tbt{call }\pn{proc$_{i+1}$}\in\ST^{\pn{proc$_{i}$}},\] for $i=0,\dots,k-1$ that does not reduce any input qubit list and such that \(\pn{proc$_0$}=\pn{proc$_k$}\).
\end{definition}

It is clear that $\wf{}\subsetneq\wf{}^\star$, as for any sequence of procedure calls that would serve as counterexample to Definition~\ref{def:weak-wf}, $\pn{proc$_0$}\sim \dots \sim \pn{proc$_{k-1}$}$. We now show that these two sets of programs are \emph{extensionally equivalent}, in the sense that a weakly well-founded program can be transformed into a semantically equivalent strongly well-founded program.

\begin{restatable}{theorem}{wfstar}
\label{thm:wfstar}
Given any $\PR\in\wf{}^\star$, there exists a $\PR'\in\wf{}$ such that $\llbracket\PR\rrbracket=\llbracket\PR'\rrbracket$. Therefore, all $\wf{}^\star$ programs terminate.
\end{restatable}

Well-foundedness of a program (weak or strong) only ensures its termination, but it still allows for infeasible programs that run in exponential time on the size of the input, for instance a program with the following structure:
\vspace{3mm}

\begin{PFOQprogram}
 \tbt{decl }\pn{double}(a)\{\\
 \quad ...\ {\color{gray} \# some operations}\\
 \quad \tbt{call }\pn{double}(a-[0]);\\
 \quad \tbt{call }\pn{double}(a-[0]);\ \}
\end{PFOQprogram}
\vspace{3mm}

To ensure a polynomial-time execution, a \emph{bounded-width} condition is added to prevent recursive duplication of procedure calls.

\paragraph*{Polynomial time.}

To keep track of the doubling of recursive procedure calls, we use the following notions of \emph{width of a procedure} and \emph{width of a program} from~\cite{HPS23}.

\begin{definition}[Width of a procedure] Given a program \PR{} and a procedure $\proc\in\PR$, the width of $\proc$ in $\PR$, denoted $\width_\PR(\proc)$, and the width of statement $\ST$ in \PR{} relative to procedure $\proc$, denoted $w^\proc{}_\PR(\ST)$, are two non-negative integers. They are defined inductively as:
\begin{align*}
\width_\PR(\proc) &\triangleq w_\PR^\proc(\ST^\proc)
\end{align*}
where
\begin{align*}
& w_\PR^\proc(\tbt{skip}) \triangleq 0,\\
& w_\PR^\proc(\textnormal{\texttt{l[i]*= U}}) \triangleq 0,\\
& w_\PR^\proc(\ST_1\ \ST_2) \triangleq w_\PR^\proc(\ST_1) + w_\PR^\proc(\ST_2)\\
& w_\PR^\proc(\tbt{if }(\textnormal{\texttt{b}})\tbt{ then }\{\ST_1\}\tbt{ else }\{\ST_2\})\\
& \qquad \triangleq \max(w_\PR^\proc(\ST_1),w_\PR^\proc(\ST_2))\\
& w_\PR^\proc(\textnormal{\texttt{\tbt{qcase }l[i]\tbt{ of }\{0 -> \ST$_0$,1 -> \ST$_1$\}}})\\
&\qquad \triangleq \max(w_\PR^\proc(\ST_0),w_\PR^\proc(\ST_1))\\
& w_\PR^\proc(\tbt{call }\pn{proc'})\triangleq 
\begin{cases}
1, & \text{if }\proc\sim_\PR\pn{proc'},\\
0 & \text{otherwise.}
\end{cases}
\end{align*}
\end{definition}

\begin{definition} The \emph{width of a program} $\PR\in\foq{}$ is defined as the maximum value of $\width_\PR(\proc)$ such that $\proc\in\PR$.
\end{definition}

Programs that avoid duplication of recursive procedure calls are defined as having \emph{bounded width}.

\begin{definition}[Bounded width] A program $\PR\in\foq{}$ is said to have bounded width, written $\PR\in \wi_{\leq 1}$, if the width of $\PR$ does not exceed 1.
\end{definition}

Putting together the well-foundedness and bounded width conditions, we obtain a polynomial-time fragment of \foq{}.

\begin{definition}[Polynomial-time programs] Let $\pfoq{}^\star$ denote the set of weakly well-founded programs with bounded width, i.e.
\[ \pfoq{}^\star \triangleq \wf{}^\star \cap \wi_{\leq 1}.\]
\end{definition}

In~\cite{HPS23}, the strongly well-founded, bounded-width fragment of \foq{}, \(\pfoq{} \triangleq \wf{} \cap \wi_{\leq 1}\) is sound and complete for quantum polynomial time. This is also true for the more expressive, weakly well-founded version.

\begin{restatable}{theorem}{characterization}
\label{thm:characterization}
$\pfoq{}^\star$ is sound and complete for quantum polynomial time.
\end{restatable}
\begin{proof}
Completeness is trivial since $\pfoq{}\subseteq \pfoq{}^\star$. Soundness is given by the extensional equivalence shown in the proof of Theorem~\ref{thm:wfstar} applied to $\pfoq{}^\star$. A full proof is given in Appendix~\ref{app:proofs}.
\end{proof}

\paragraph*{Polylogarithmic time.}

We now consider the \emph{polylogarithmic-time} fragment of \foq{}, defined by replacing well-foundedness by a strictly stronger \emph{recursively halving} condition.

\begin{definition}[Strong recursive halving, extended from~{\cite[Definition 3]{FHPS25}}]
A \foq{} program \PR{} is said to be \emph{strongly recursively halving}, written $\PR\in\half$ if, for every pair of procedures $\proc,\, \pn{proc'} \in\PR$ such that $\proc\sim_\PR \pn{proc'}$, any procedure call to $\pn{proc'}$ appearing in $\textnormal{\ST}^\proc$ reduces the number of elements in some input qubit list by at least half.
\end{definition}

The fragment $\lfoq{}\triangleq \half{}\cap \wi_{\leq 1}$ captures polylogarithmic time for quantum random-access Turing machines~\cite{FHPS25}.
 
Similarly to $\wf{}^\star$ (Definition~\ref{def:weak-wf}), the weak version of recursive halving is given by considering a reduction of input qubit lists by half over any call cycle in a program.

\begin{definition}[Weakly recursive halving]\label{def:weak-rec-halv}
A program is said to be \emph{weakly recusively-halving}, written $\PR\in\half^\star$, if there is no sequence of procedure calls
\[\tbt{call }\pn{proc$_{i+1}$}\in\ST^{\pn{proc$_{i}$}},\] for $i=0,\dots,k-1$ that do not reduce any input qubit list \emph{by half} and such that \(\pn{proc$_0$}=\pn{proc$_k$}\).
\end{definition}

Letting $\lfoq^\star \triangleq \half{}^\star \cap \wi_{\leq 1}$ we show the following result.

\begin{restatable}{theorem}{logtimecharacterization}
\label{thm:logtime-characterization}
$\lfoq{}^\star$ is sound and complete for quantum polylogarithmic time.
\end{restatable}
\begin{proof}
The argument is analogous to the proof of Theorem~\ref{thm:characterization}.
\end{proof}

Notice that by construction $\lfoq{}^\star \subsetneq \pfoq{}^\star$. We now consider the compilation strategy of~\cite{HPS25} applied to both $\lfoq{}^\star$ and $\pfoq{}^\star$. The compilation by the \pfoqcompiler{} is denoted \[\tb{compile}:\text{Programs}\times \text{List}(\mathbb{N})\mapsto\text{Circuits},\]
that is, a function from programs and input sizes into circuits. We can now give bounds on the circuits obtained for the weaker fragments $\lfoq^\star$ and $\pfoq{}^\star$. 

\begin{restatable}{theorem}{compilation}
\label{thm:compilation}
Given a program $\PR\in\foq{}$, let $C_\mathcal{N}\triangleq \tb{compile}(\PR,\mathcal{N})$. Then
\begin{itemize}
\item if $\PR\in\pfoq{}^\star$, then $\textnormal{size}(C_\mathcal{N})=\mathcal{O}(\textnormal{poly}(\mathcal{N}))$,
\item if $\PR\in\lfoq{}^\star$, then $\textnormal{size}(C_\mathcal{N})=\mathcal{O}(\textnormal{poly}(\mathcal{N}))$ and \textnormal{depth}$(C_\mathcal{N})=\mathcal{O}(\textnormal{polylog}(\mathcal{N}))$.
\end{itemize}
\end{restatable}
\begin{proof} This follows from the extensional equivalence between $\lfoq{}$ and $\lfoq{}^\star$ and $\pfoq{}$ and $\pfoq{}^\star$ and the compilation results of~\cite{HPS23} and~\cite{FHPS25}.
\end{proof}

We now turn to the relation between the circuit complexity of a \foq{} program $\PR$ and its time complexity in the operational semantics, given by the Time$_\PR$ function described in Section~\ref{ss:semantics}.

\paragraph*{Eliminating the cost of branch sequentialization.}

The overhead cost of merging different procedure calls is given by the extra ancillae and gates necessary to anchor the procedures and control-swap the different qubit addresses. This cost can be avoided if the program is structured such that merging has $\mathcal{O}(1)$ cost, for instance, by ensuring that the inputs of merged procedures already correspond exactly, and therefore do not need to be rewired.

Furthermore, even in the context of constant-overhead merging, the cost of branch sequentialization can arise from procedure calls being incompatible due to mismatched classical inputs or quantum input sizes. For this reason, one may impose further restrictions on programs that ensure \emph{uniformity} over procedure calls, as in the following definition.

\begin{definition}[Strong uniformity, adapted from~\cite{HPS25}] \label{def:strong-unif}
A program is said to be \emph{strongly uniform} if:
\begin{enumerate}[label= (\roman*)]
\item procedures make no use of integer variables;\label{stunif-i}
\item procedures only have one qubit parameter;\label{stunif-ii}
\item qubit parameters are passed as pure variables or some fixed reduction strategy in procedure calls.\label{stunif-iii}
\end{enumerate}
\end{definition}

The strong uniformity conditions are relatively easy to check. For instance, condition~\ref{stunif-iii} states that in all procedure calls of the form
$\tbt{call }\proc(\texttt{l});$ (notice that by conditions~\ref{stunif-i} and~\ref{stunif-ii} this is the most general case) we have that either $\texttt{l}=\texttt{a}$ for some qubit list variable $\texttt{a}$ or there is some fixed expression, such as $\texttt{a-[1]}$, that is used throughout the program for all procedure calls.

For example, program \texttt{\color{gray}QFT} in Section~\ref{ss:parametrized-gates} is not \emph{strongly} uniform since the procedure calls $\tbt{call }\pn{qft}\texttt{(a-[0])}$ and $\tbt{call }\pn{rot}\texttt{(a-[1])}$ both appear in \texttt{\color{gray}QFT}.

Uniform programs give the best bound on compilation (\cite{HPS25}, similar to the upcoming Theorem~\ref{thm:weak-unif-circ}). However, the strong uniformity condition is quite strong, so it is interesting from the point of view of expressivity to relax this condition somewhat while preserving the same compilation bounds.

One motivation for the restriction is that we want merged procedure calls to have exactly the same input qubits, so that no rearranging of qubit addresses needs to be done and therefore merging can be done in constant time (as in the \texttt{\color{gray}PAIRS} example in Section~\ref{s:introduction}). The same result can be achieved if we allow for two additional strategies, here defined as \emph{front-reducing} and \emph{back-reducing}. We collectively call these strategies, together with strong uniformity, \emph{similarly reducing}.

\begin{definition}[Similarly reducing] A \emph{similarly reducing} strategy is one of the following
\begin{itemize}
\item \textbf{strong uniformity:} either $\textnormal{\texttt{l}}=\textnormal{\texttt{q}}$ or $\texttt{l}=\textnormal{\texttt{l}}_0$, for some fixed expression $\textnormal{\texttt{l}}_0$
\item \textbf{front-reducing:}
\[ \textnormal{ $\texttt{fr}:=\texttt{a}\mid \texttt{(fr)$^+$} \mid \texttt{fr - [0,1,\dots,$n$]}$} \]
\item \textbf{back-reducing:} \[ \textnormal{ $\texttt{br}:=\texttt{a}\mid \texttt{(br)$^-$} \mid \texttt{br - [-1,\dots,$-n$]}$} \]
\end{itemize}
\end{definition}

We can now define the \emph{weak} version of the uniformity condition.

\begin{definition}[Weak uniformity]\label{def:weak-uniformity}
A \foq{} program $\PR$ is said to be \emph{weakly uniform}, written $\PR\in\unif{}^\star$, if $\PR$ satisfies the following restrictions:
\begin{enumerate}[label= (\roman*)]
\item procedures make no use of integer variables;\label{unif-i}
\item procedures only have one qubit parameter;\label{unif-ii}
\item all (mutually) recursive procedure calls and any procedure calls occurring in branches that are orthogonal to the recursive branches are \emph{similarly reducing}.\label{unif-iii}
\end{enumerate}  
\end{definition}
~
In the above definition, we have replaced the fixed reduction strategy of Definition~\ref{def:strong-unif} with the more general concept of similarly reducing strategies, and this makes the \QFT{} program weakly uniform (whereas it was \emph{not} strongly uniform).

Another way that this increases expressivity is that the check is no longer performed on \emph{all} procedure calls but rather those performed between mutually recursive procedures (that is to say, those performed in recursive branches) and also those performed orthogonally to the recursive branches.

For instance, in the following procedure (let \pn{g} be defined such that $\pn{f}\succ \pn{g}$):

\noindent
\begin{minipage}{\columnwidth}
\centering
\scalebox{0.9}{
\begin{PFOQprogram}
\tbt{decl }\pn{f}(\texttt{a})\{\\
\quad \tbt{call }\pn{g}(a-[1]);\\
\quad \tbt{if }|a|>1\tbt{ then }\{\ \tbt{call }\pn{f}(a-[0]);\ \}\\
\quad \tbt{else }\{ \tbt{call }\pn{g}(a);\ \}\ \}
\end{PFOQprogram}}
\vspace{3mm}
\end{minipage}
There is only one recursive branch, containing a call to \pn{g} on input \texttt{a-[1]} and a call to \pn{f} on input \texttt{a-[0]}. Of these two, only the call to \pn{f} is recursive. The call orthogonal to this is the call to \pn{g} on input \texttt{a}. Therefore, we need only check the calls to \pn{f} and the second call to \pn{g} to conclude that \pn{f} is back-reducing.

Weak uniformity is strictly less restrictive than strong uniformity, but we show that it does not change the expressivity of the language.

\begin{restatable}{theorem}{unifcharacterization}
\label{thm:unf-characterization}
$\pfoq{}^\star\cap \unif{}^\star$ is sound and complete for quantum polynomial time.
\end{restatable}
\begin{proof}
In Appendix~\ref{app:proofs}.
\end{proof}

In the weakly uniform fragment, we can show that the complexity of compiling the \tbt{qcase} statement is the maximum of the branches, which is shown by the following circuit bound.

\begin{restatable}{theorem}{weakunifcirc}
\label{thm:weak-unif-circ}
Let $\PR\in\pfoq{}^\star \cap \unif{}^\star$, then \[\textnormal{size}(\tb{compile}(\PR,\mathcal{N}))=\mathcal{O}(\textnormal{Time}_{\PR}(\mathcal{N})).\]
\end{restatable}
\begin{proof}
In Appendix~\ref{app:proofs}.
\end{proof}

Having shown the properties of different fragments, we now consider their static verification.

\begin{table*}[t]
\caption{Compilation strategies used in different languages and the circuit complexity of quantum control.}
\centering
\scalebox{0.95}{
\begin{tabular}{|Sc|Sc|Sc|Sc|Sc|}
\multicolumn{1}{c}{} & \multicolumn{3}{c}{\begin{tabular}{c}
\textbf{Circuit complexity of quantum control}\\[2mm]
\end{tabular}} & \multicolumn{1}{c}{}\\
\cline{2-4}
\multicolumn{1}{c|}{\textbf{Strategy}} & \textbf{width} & \textbf{depth} & \textbf{size} & \multicolumn{1}{|c}{\textbf{Languages}} \\
\cline{1-5}
sequential & \maxcolor{$\mathcal{O}(\max(W_1,W_2))$} & $\mathcal{O}(D_1+D_2)$ & $\mathcal{O}(S_1+S_2)$ & Proto-Quipper, Tower \\
parallel & $\mathcal{O}^\text{poly}(W_1+W_2)$ & \maxcolor{${\mathcal{O}}^\text{poly}(\max(D_1,D_2))$}& ${\mathcal{O}}^\text{poly}(S_1+S_2)$ & RQC$^{++}$, Qrisp
\\
merging & \maxcolor{${\mathcal{O}}^\text{poly}(\max(W_1,W_2))$}  & \maxcolor{$\tilde{\mathcal{O}}(\max(D_1,D_2))$} &  \maxcolor{${\mathcal{O}}^\text{poly}(\max(S_1,S_2))$} & \pfoq{}$^\star$ \\
merging & \maxcolor{$\mathcal{O}(\max(W_1,W_2))$}  & \maxcolor{$\mathcal{O}(\max(D_1,D_2))$} &  \maxcolor{$\mathcal{O}(\max(S_1,S_2))$} & \pfoq{}$^\star \cap \unif{}^\star$ \\
\cline{1-5}\end{tabular}}
\label{table:compilation-strategies}
\end{table*}

\paragraph*{Static verification.} The \pfoqcompiler{} statically verifies whether a program is weakly well-founded, weakly recursively halving or uniform. This static verification can be done efficiently on the program size.

\begin{restatable}[Static analysis]{prop}{staticanalysis}
\label{prop:staticanalysis}
Given a program $\PR\in\foq{}$, the properties
\[\PR\in \{\wf{}^\star,\, \half^\star,\, \wi_{\leq 1},\,\unif^\star\}\]
can be checked in time $O(|\PR|^2)$.
\end{restatable}
\begin{proof}
In Appendix~\ref{app:proofs}.
\end{proof}

The static analysis information can be accessed using the \texttt{--verbose} flag. The compiler then outputs information on whether the program belongs to the $\wf{}^\star$, $\half{}^\ast$, $\wi_{\leq 1}$, and $\unif{}^\ast$ fragments, with the number of ancillae required for the compilation, as well as the total circuit size and depth.

Whenever the program does \emph{not} satisfy a given condition, the compiler gives an example of the unmet condition. For instance, the program \texttt{\color{gray} PAIRS} run on input size 8 will result in the following output:\\
\vspace{-2mm}

\noindent
\texttt{Properties:}\\
\texttt{- Well-founded}\\
\texttt{- NOT halving. An example call cycle that does not reduce qubits by half: pairs -> pairs}\\
\texttt{- Bounded width: 1}\\
\texttt{- Uniform}\\
\\
\texttt{Compiled circuit using 3 ancillae, 12 total gates and depth 12.}

\vspace{2mm}
The program is weakly well-founded, as \pn{pairs} always reduces its input list \texttt{a} by at least one qubit (in this case, two) for both recursive calls. However, since this does not reduce the number of qubits by half, the program is not recursively halving.

\section{Compilation example}
\label{s:compilation-example}

The reader interested in the details of the compilation strategy is invited to read~\cite{HPS25} for the theoretical background of this work. By way of illustration, we give here the steps of a more involved compilation of a \pfoq{} program.

In Section~\ref{s:introduction} we described the compilation of the \texttt{\color{gray}PAIRS} program using the merging strategy. The simplicity of the \pn{pairs} procedure allowed for a very straightforward combination of the two recursive calls into a single one. However, if the two procedure calls differ even in a slight way (making them not compatible, in the sense described in Section~\ref{sec:restrictions}), then the compilation can no longer be done directly with a simple rule as given in Figure~\ref{fig:merging-strategy}.

To better illustrate the compilation strategy, we introduce a program \texttt{\color{gray}PEARS} that differs from \texttt{\color{gray}PAIRS} only very slightly: the procedure calls now differ in their register, and the guard of the if-statement has also been changed.

Because the two recursive calls in \pn{pears} are no longer the same, they cannot be simplified immediately. Rather, we must run the compilation, identifying compatible instances of \pn{pears} and merging them on the fly.

\vspace{3mm}

\noindent
\begin{minipage}{0.45\columnwidth}
\scalebox{0.85}{
\begin{PFOQprogram}
{\color{gray}PEARS}\\
\\
\tbt{decl }\pn{pears}(a)\{\\
\quad \tbt{if }(|a|{\color{purple}>2})\tbt{ then }\{ {\color{gray} \# recursive case}\\
\qquad  \tbt{qcase }(a[0],a[1])\tbt{ of }\{ \\
\qquad\quad  00 -> \tbt{call }\pn{pears}(a-[0,1{\color{purple},2}]);,\\
 \qquad\quad 01 -> \tbt{skip};,\\
 \qquad\quad 10 -> \tbt{skip};,\\
\qquad\quad  11  -> \tbt{call }\pn{pears}(a-[0,1]);\ \}\\
\quad  \} \tbt{else} \{ a[0] *= NOT; \}\ \} {\color{gray} \# base case}\\
::\\
\tbt{define }q;\\
::\\
\tbt{call }\pn{pears}(q);
\end{PFOQprogram}}
\end{minipage}
\vspace{3mm}

\begin{figure*}
\centering
\begin{tabular*}{\textwidth}{cccccccc}
$\qquad\quad$
\scalebox{0.85}
{
\begin{quantikz}[wire types = {b},classical gap=0.07cm,row sep = {4mm, between origins}]
& \gate[style={fill=gray!50}]{}\compgroup{1}{1} &
\end{quantikz}}
&
$\quad\rightarrow^{a}\quad$
&
\scalebox{0.85}
{
\begin{quantikz}[wire types = {b},classical gap=0.07cm,row sep = {4mm, between origins}]
& \gate{}\compgroup{1}{1} &
\end{quantikz}}
&
$\quad\rightarrow^{d}\quad$
&
\scalebox{0.7}
{\begin{quantikz}[wire types = {q,q,q,b},classical gap=0.07cm, row sep = {4mm, between origins}]
& \octrl{1}\compgroup{4}{2} & \ctrl{1} &\\
& \octrl{2} & \ctrl{1} &\\
& 					 & \gate[2,style={fill=gray!50}]{} &\\
& \gate[style={fill=gray!50}]{} & &
\end{quantikz}}
&
$\quad\rightarrow^{a}\quad$
&
\scalebox{0.7}
{
\begin{quantikz}[wire types = {q,q,q,b,q},classical gap=0.07cm, row sep = {4mm, between origins}, column sep = 1mm]
& \ctrl{1} & [5pt]\octrl{1}\compgroup{5}{2}& & [5pt]\ctrl{1} &\\
& \ctrl{3} & \octrl{2} & & \ctrl{3} &\\
& 					 &  &\gate[2]{} & &\\
&  & \gate[style={fill=gray!50}]{} & & &\\
\lstick{$a_5$} & \targ{} & & \ctrl{-1} & \targ{} &
\end{quantikz}
}
&
\end{tabular*}\\
\vspace{3mm}
\noindent
\begin{tabular*}{\textwidth}{ccccc}
& & & & $\downarrow_{d}$\\
\scalebox{0.75}{
\begin{quantikz}[wire types = {q,q,q,q,q,q,q,q,q},classical gap=0.07cm, row sep = {4mm, between origins}, column sep = 0.9mm]
& \ctrl{1} & \octrl{1}& [5pt]\compgroup{9}{4} & && &[5pt]\octrl{1} &\ctrl{1} &\\
& \ctrl{6} & \octrl{7} & & && &\octrl{7}&\ctrl{6} &\\
& & 					 &  &&\octrl{1} &\ctrl{1}& & &\\
&&  & \octrl{1} & \ctrl{1} &\octrl{2} & \ctrl{1}& & &\\
& & & \octrl{2} & \ctrl{1} && \gate[3,style={fill=gray!50}]{}& & &\\
& & & &\gate[2,style={fill=gray!50}]{} &\gate[2,style={fill=gray!50}]{}& & & &\\
& & &\gate[,style={fill=gray!50}]{} & & & & & & \\
\lstick{$a_5$} & \targ{} & & &&\ctrl{-1} & \ctrl{-1} & &\targ{} & \\
\lstick{$a_4$} & &\targ{} & \ctrl{-2} &\ctrl{-2}& & &\targ{}& &
\end{quantikz}}
& 
$\leftarrow^d$
& 
{\scalebox{0.8}{
\begin{quantikz}[wire types = {q,q,q,q,q,b,q,q},classical gap=0.07cm, row sep = {4mm, between origins}, column sep = 0.7mm]
& \ctrl{1} & \octrl{1}& [5pt]\compgroup{8}{3}& & &[5pt]\octrl{1} &\ctrl{1} &\\
& \ctrl{5} & \octrl{6} & & & &\octrl{6}&\ctrl{5} &\\
& & 					 &  &\octrl{1} &\ctrl{1}& & &\\
&&  & \gate[3]{} & \octrl{2} & \ctrl{1}& & &\\
& & & &  & \gate[2,style={fill=gray!50}]{}& & &\\
& & & & \gate[,style={fill=gray!50}]{}& & & &\\
\lstick{$a_5$} & \targ{} & & &\ctrl{-1} & \ctrl{-1} & &\targ{} &\\
\lstick{$a_4$} & &\targ{} & \ctrl{-2} & & &\targ{}& &
\end{quantikz}}}
& 
$\leftarrow^a$
&
\scalebox{0.7}
{\begin{quantikz}[wire types = {q,q,q,q,q,b,q},classical gap=0.07cm, row sep = {4mm, between origins}, column sep = 1mm]
& \ctrl{1} & [5pt]\octrl{1}\compgroup{7}{3}& & &[5pt]\ctrl{1} &\\
& \ctrl{5} & \octrl{2} & & &\ctrl{5} &\\
& 					 &  &\octrl{1} &\ctrl{1} & &\\
&  & \gate[3,style={fill=gray!50}]{} & \octrl{2} & \ctrl{1} & &\\
& & &  & \gate[2,style={fill=gray!50}]{} & &\\
& & & \gate[,style={fill=gray!50}]{} & & &\\
\lstick{$a_5$} & \targ{} & & \ctrl{-1} & \ctrl{-1} & \targ{} &
\end{quantikz}}\\
&&&&\\[-2mm]
$\downarrow_a$&&&&\\[-4mm]
&&&&\\
\scalebox{0.65}{
\begin{quantikz}[wire types = {q,q,q,q,q,q,q,q,q,q},classical gap=0.07cm, row sep = {4mm, between origins}, column sep = 1mm]
& \ctrl{1} & \octrl{1}& &[5pt]\compgroup{10}{4} && &&[5pt]&\octrl{1} &\ctrl{1} &\\
& \ctrl{6} & \octrl{7} & & && &&&\octrl{7}&\ctrl{6} &\\
& & 					 &  \ctrl{1} &&&\octrl{1} & &\ctrl{1} & &&\\
&&  &\ctrl{4} &\octrl{1} & \ctrl{1} &\octrl{2} & & \ctrl{4}& &&\\
& & &  &\octrl{2} & \ctrl{1} && \gate[3]{}& & &&\\
& & & &&\gate[2,style={fill=gray!50}]{} &\gate[2,style={fill=gray!50}]{}&& & & &\\
& & &&\gate[,style={fill=gray!50}]{} & & & & & & &\\
\lstick{$a_5$} & \targ{} & & \ctrl{2}& &&\ctrl{-1} &   & \ctrl{2}& &\targ{} & \\
\lstick{$a_4$} & &\targ{} & &\ctrl{-3} &\ctrl{-3}& & &&\targ{}& &\\
\lstick{$a_3$} & &  &\targ{} &  & & &\ctrl{-3} & \targ{} && &
\end{quantikz}}
&
$\rightarrow^d$
&
\scalebox{0.65}{
\begin{quantikz}[wire types = {q,q,q,q,q,q,q,q,q,q},classical gap=0.07cm, row sep = {4mm, between origins}, column sep = 0.7mm]
& \ctrl{1} & \octrl{1}& &[5pt]\compgroup{10}{4} && &&[5pt]&\octrl{1} &\ctrl{1} &\\
& \ctrl{6} & \octrl{7} & & && &&&\octrl{7}&\ctrl{6} &\\
& & 					 &  \ctrl{1} &&&\octrl{1} & &\ctrl{1} & &&\\
&&  &\ctrl{4} &\octrl{1} & \ctrl{1} &\octrl{2} & & \ctrl{4}& &&\\
& & &  &\octrl{2} & \ctrl{1} && \ctrl{1}& & &&\\
& & & &&\gate[2,style={fill=gray!50}]{} &\gate[2,style={fill=gray!50}]{}&\ctrl{1}& & & &\\
& & &&\gate[,style={fill=gray!50}]{} & & & \gate[,style={fill=gray!50}]{}& & & &\\
\lstick{$a_5$} & \targ{} & & \ctrl{2}& &&\ctrl{-1} &   & \ctrl{2}& &\targ{} & \\
\lstick{$a_4$} & &\targ{} & &\ctrl{-3} &\ctrl{-3}& & &&\targ{}& &\\
\lstick{$a_3$} & &  &\targ{} &  & & &\ctrl{-3} & \targ{} && &
\end{quantikz}}
&
$\rightarrow^a$
&
\hspace{-7mm}
\scalebox{0.65}{
\begin{quantikz}[wire types = {q,q,q,q,q,q,q,q,q,q},classical gap=0.07cm, row sep = {4mm, between origins}, column sep = 0.7mm]
& \ctrl{1} & \octrl{1}& & &[5pt]\compgroup{11}{4}& &&&[5pt]&&\octrl{1} &\ctrl{1} &\\
& \ctrl{6} & \octrl{7} & & && &&&&&\octrl{7}&\ctrl{6} &\\
& & 					 &  \ctrl{1} &&&&\octrl{1} & &&\ctrl{1} & &&\\
&&  &\ctrl{4} &\ctrl{1} &\octrl{1} &&\octrl{2} & &\ctrl{1} &\ctrl{4}& &&\\
& & &  &\ctrl{4} & \octrl{2} &&& \ctrl{1}&\ctrl{4}& & &&\\
& & & &&&\gate[2]{} &\gate[2,style={fill=gray!50}]{}&\ctrl{1}& && & &\\
& & && &\gate[,style={fill=gray!50}]{} && & \gate[,style={fill=gray!50}]{}&& & & &\\
\lstick{$a_5$} & \targ{} & & \ctrl{2}& &&&\ctrl{-1} & && \ctrl{2}& &\targ{} & \\
\lstick{$a_4$} & &\targ{} & &\ctrl{2} &\ctrl{-2} && &&\ctrl{2} &&\targ{}& &\\
\lstick{$a_3$} & &  &\targ{} &  & && &\ctrl{-3} && \targ{} && &\\
\lstick{$a_2$} & &  & &\targ{}  &&\ctrl{-4} & & & \targ{}& && &
\end{quantikz}}\\
&&&&\\[-2mm]
&&&&$\downarrow_m$\\[-4mm]
&&&&\\
\scalebox{0.65}{
\begin{quantikz}[wire types = {q,q,q,q,q,q,q,q,q,q},classical gap=0.07cm, row sep = {4mm, between origins}, column sep = 0.7mm]
& \ctrl{1} & \octrl{1}& & &&& &[5pt]\compgroup{12}{2}&&[5pt]&&&&\octrl{1} &\ctrl{1} &\\
& \ctrl{6} & \octrl{7} & & &&& &&&&&&&\octrl{7}&\ctrl{6} &\\
& & 					 &  \ctrl{1} &&\octrl{1} &&&&&&\octrl{1} &&\ctrl{1} & &&\\
&&  &\ctrl{4} &\ctrl{1} &\octrl{4}&&\octrl{1}& & &\octrl{1}&\octrl{4} &\ctrl{1} &\ctrl{4}& &&\\
& & &  &\ctrl{4} & &&\octrl{4} &&\ctrl{1}&\octrl{4} &&\ctrl{4}& & &&\\
& & & &&& &&&\ctrl{1}& && && & &\\
& & &&&& \targ{} & &\gate{}&\gate[,style={fill=gray!50}]{} & &&& & & &\\
\lstick{$a_5$} & \targ{} & & \ctrl{2}&&\ctrl{3} &&&&& &\ctrl{3}  && \ctrl{2}& &\targ{} & \\
\lstick{$a_4$} & &\targ{} & &\ctrl{2} &&&\ctrl{3} && &\ctrl{3}&&\ctrl{2} &&\targ{}& &\\
\lstick{$a_3$} & &  &\targ{} &  && &&&\ctrl{-3} && & &\targ{} && &\\
\lstick{$a_2$} & &  & &\targ{}  &\targ{}&\ctrl{-4}&&& & &\targ{} & \targ{}& && &\\
\lstick{$a_1$} & &  & &  & & &\targ{}& \ctrl{-5}  & &\targ{} &  & && &&
\end{quantikz}}
&
$\leftarrow^a$
&
\hspace{-10mm}
\scalebox{0.65}{
\begin{quantikz}[wire types = {q,q,q,q,q,q,q,q,q,q},classical gap=0.07cm, row sep = {4mm, between origins}, column sep = 0.8mm]
& \ctrl{1} & \octrl{1}& & &&&[5pt]\compgroup{11}{2} &&[5pt]&&&\octrl{1} &\ctrl{1} &\\
& \ctrl{6} & \octrl{7} & & &&& &&&&&\octrl{7}&\ctrl{6} &\\
& & 					 &  \ctrl{1} &&\octrl{1} &&&&\octrl{1} &&\ctrl{1} & &&\\
&&  &\ctrl{4} &\ctrl{1} &\octrl{4}&&\octrl{1} & &\octrl{4} &\ctrl{1} &\ctrl{4}& &&\\
& & &  &\ctrl{4} & &&\octrl{2} &\ctrl{1}& &\ctrl{4}& & &&\\
& & & &&& &&\ctrl{1} && && & &\\
& & &&&& \targ{} &\gate[,style={fill=gray!50}]{} &\gate[,style={fill=gray!50}]{} & && & & &\\
\lstick{$a_5$} & \targ{} & & \ctrl{2}&&\ctrl{3} &&& &\ctrl{3}  && \ctrl{2}& &\targ{} & \\
\lstick{$a_4$} & &\targ{} & &\ctrl{2} &&&\ctrl{-2} & &&\ctrl{2} &&\targ{}& &\\
\lstick{$a_3$} & &  &\targ{} &  && && \ctrl{-3}&& & \targ{} && &\\
\lstick{$a_2$} & &  & &\targ{}  &\targ{}&\ctrl{-4}&&  &\targ{} & \targ{}& && &
\end{quantikz}}
\hspace{-7mm}
&
$\leftarrow^d$
&
\hspace{-7mm}
\scalebox{0.65}{
\begin{quantikz}[wire types = {q,q,q,q,q,q,q,q,q,q},classical gap=0.07cm, row sep = {4mm, between origins}, column sep = 0.8mm]
& \ctrl{1} & \octrl{1}& & &&[5pt]\compgroup{11}{3}& &&[5pt]&&&\octrl{1} &\ctrl{1} &\\
& \ctrl{6} & \octrl{7} & & &&& &&&&&\octrl{7}&\ctrl{6} &\\
& & 					 &  \ctrl{1} &&\octrl{1} &&&&\octrl{1} &&\ctrl{1} & &&\\
&&  &\ctrl{4} &\ctrl{1} &\octrl{4}&\octrl{1} && &\octrl{4} &\ctrl{1} &\ctrl{4}& &&\\
& & &  &\ctrl{4} & &\octrl{2} &&\ctrl{1}& &\ctrl{4}& & &&\\
& & & &&& &\gate[2]{}&\ctrl{1} && && & &\\
& & &&&& \gate[,style={fill=gray!50}]{}& &\gate[,style={fill=gray!50}]{} & && & & &\\
\lstick{$a_5$} & \targ{} & & \ctrl{2}&&\ctrl{3} &&& &\ctrl{3}  && \ctrl{2}& &\targ{} & \\
\lstick{$a_4$} & &\targ{} & &\ctrl{2} &&\ctrl{-2} && &&\ctrl{2} &&\targ{}& &\\
\lstick{$a_3$} & &  &\targ{} &  && && \ctrl{-3}& && \targ{} && &\\
\lstick{$a_2$} & &  & &\targ{}  &\targ{}&&\ctrl{-4} & &\targ{} & \targ{}& && &
\end{quantikz}}\\
&&&&\\[-2mm]
$\downarrow_m$ &&&&\\[-2mm]
&&&&\\
\hspace{-7mm}
\scalebox{0.65}{
\begin{quantikz}[wire types = {q,q,q,q,q,q,q,q,q,q},classical gap=0.07cm, row sep = {4mm, between origins}, column sep = 0.7mm]
& \ctrl{1} & \octrl{1}& & &&& & &[5pt]\compgroup{12}{1}&[5pt]&&&&&\octrl{1} &\ctrl{1} &\\
& \ctrl{6} & \octrl{7} & & &&& &&&&&&&&\octrl{7}&\ctrl{6} &\\
& & 					 &  \ctrl{1} &&\octrl{1} &&&&&&&\octrl{1} &&\ctrl{1} & &&\\
&&  &\ctrl{4} &\ctrl{1} &\octrl{4}&&\octrl{1}&& & &\octrl{1}&\octrl{4} &\ctrl{1} &\ctrl{4}& &&\\
& & &  &\ctrl{4} & &&\octrl{4} &\ctrl{1}&&\ctrl{1}&\octrl{4} &&\ctrl{4}& & &&\\
& & & &&& &&\ctrl{4}&&\ctrl{4} &&& && & &\\
& & &&&& \targ{} & & &\gate[]{}& & &&& & & &\\
\lstick{$a_5$} & \targ{} & & \ctrl{2}&&\ctrl{3} &&&&&&&\ctrl{3}  && \ctrl{2}& &\targ{} & \\
\lstick{$a_4$} & &\targ{} & &\ctrl{2} &&&\ctrl{3} && &&\ctrl{3}&&\ctrl{2} &&\targ{}& &\\
\lstick{$a_3$} & &  &\targ{} &  && &&\ctrl{2}& &\ctrl{2}& && &\targ{} && &\\
\lstick{$a_2$} & &  & &\targ{}  &\targ{}&\ctrl{-4}&&&&& &\targ{} & \targ{}& && &\\
\lstick{$a_1$} & &  & &  & & &\targ{}& \targ{}  &\ctrl{-5}&\targ{}&\targ{} &  & && &&
\end{quantikz}}
&\hspace{-7mm}
$\rightarrow^d$
&\hspace{-10mm}
\scalebox{0.65}{
\begin{quantikz}[wire types = {q,q,q,q,q,q,q,q,q,q},classical gap=0.07cm, row sep = {4mm, between origins}, column sep = 0.7mm]
& \ctrl{1} & \octrl{1}& & &&& & &&&&&&&\octrl{1} &\ctrl{1} &\\
& \ctrl{6} & \octrl{7} & & &&& &&&&&&&&\octrl{7}&\ctrl{6} &\\
& & 					 &  \ctrl{1} &&\octrl{1} &&&&&&&\octrl{1} &&\ctrl{1} & &&\\
&&  &\ctrl{4} &\ctrl{1} &\octrl{4}&&\octrl{1}&& & &\octrl{1}&\octrl{4} &\ctrl{1} &\ctrl{4}& &&\\
& & &  &\ctrl{4} & &&\octrl{4} &\ctrl{1}&&\ctrl{1}&\octrl{4} &&\ctrl{4}& & &&\\
& & & &&& &&\ctrl{4}&&\ctrl{4} &&& && & &\\
& & &&&& \targ{} & & &\targ{}& & &&& & & &\\
\lstick{$a_5$} & \targ{} & & \ctrl{2}&&\ctrl{3} &&&&&&&\ctrl{3}  && \ctrl{2}& &\targ{} & \\
\lstick{$a_4$} & &\targ{} & &\ctrl{2} &&&\ctrl{3} && &&\ctrl{3}&&\ctrl{2} &&\targ{}& &\\
\lstick{$a_3$} & &  &\targ{} &  && &&\ctrl{2}& &\ctrl{2}& && &\targ{} && &\\
\lstick{$a_2$} & &  & &\targ{}  &\targ{}&\ctrl{-4}&&&&& &\targ{} & \targ{}& && &\\
\lstick{$a_1$} & &  & &  & & &\targ{}& \targ{}  &\ctrl{-5}&\targ{}&\targ{} &  & && &&
\end{quantikz}}
&&
\end{tabular*}
\caption{Compilation steps for \texttt{\color{gray}PEARS} on input size 7.}
\label{fig:pears-compilation}
\end{figure*}
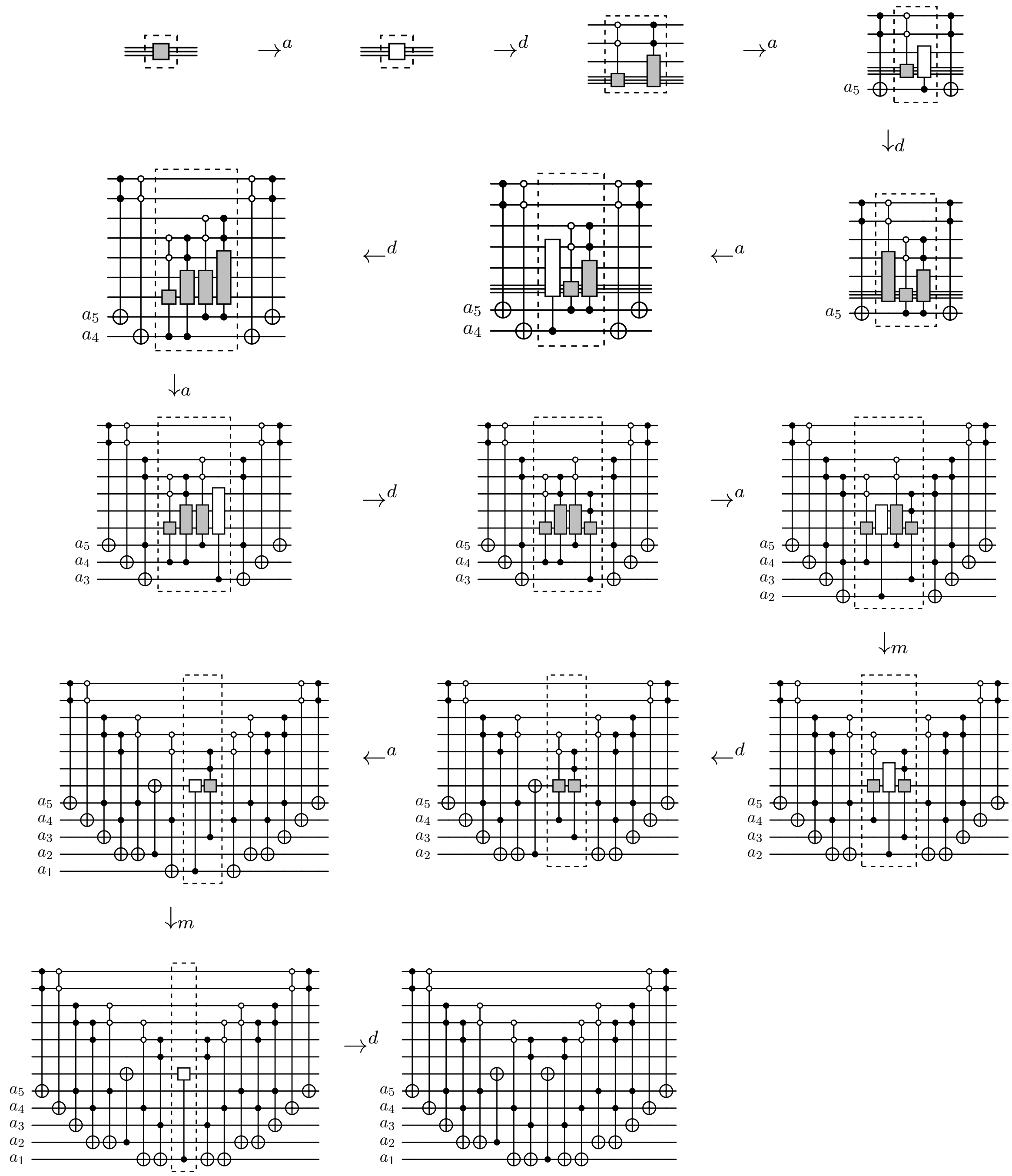

We show the steps for compiling \texttt{\color{gray}PEARS} on 7 qubits in Figure~\ref{fig:pears-compilation}. We use a white gate $\begin{quantikz} \gate{} \end{quantikz} $ to denote a circuit corresponding to $\ST^\pn{pears}$ (that is, the procedure statement of \pn{pears}) and we use a gray gate $\begin{quantikz} \gate[,style={fill=gray!50}]{} \end{quantikz} $ to represent the procedure call \tbt{call }\pn{pears}. The respective inputs are given by the input wires of the gates, and the dashed rectangle denotes the part of the circuit where the compilation is still taking place.

The compilation can be broken down into three types of steps. In the following diagrams we denote by $cs$ a quantum \emph{control structure} (that is to say, some sequence of positive and negative controls) applied to some gate or \foq{} statement being compiled:
\begin{itemize}
\item \emph{anchoring}: where a procedure call to \pn{pears} is replaced by the procedure statement $\ST^\pn{pears}$. If there is a control associated with the procedure call, then it is \emph{anchored} to an ancilla. If the size of the input is $k$, then the rule is:
\end{itemize}
\[
\scalebox{0.9}{
\begin{quantikz}[wire types = {b,b},classical gap=0.07cm, column sep = {1mm}]
&&\ \cdots\  \compgroup{2}{3} & \gate{cs}\vqw{1} &\ \cdots\ & & \\
& &\ \cdots\ & \gate[style={fill=gray!50}]{} &\ \cdots\  & &
\end{quantikz}}
\rightarrow^a
\scalebox{0.9}{
\begin{quantikz}[wire types = {b,b},classical gap=0.07cm, column sep = {1mm}]
&\gate{cs}\vqw{2} &[1mm]&\ \cdots\  \compgroup{3}{3} & &\ \cdots\ &[2mm] \gate{cs}\vqw{2} &\\
& &&\ \cdots\ & \gate[]{} &\ \cdots\  & &\\
\lstick{$a_k$}&\targ{}&&&\ctrl{-1}&&\targ{}&
\end{quantikz}}
\]
\begin{itemize}
\item \emph{merging}: if the call to \pn{pears} occurs with an input size (say, $k$), for which an instance of \pn{pears} has already been anchored, then we simply send the control to the ancilla:
\end{itemize}
\[
\scalebox{0.9}{
\begin{quantikz}[wire types = {b,b},classical gap=0.07cm, column sep = {1mm}]
&&\ \cdots\  \compgroup{3}{3} & \gate{cs}\vqw{1} &\ \cdots\ & & \\
& &\ \cdots\ & \gate[style={fill=gray!50}]{} &\ \cdots\  & &\\
\lstick{$a_k$} &&\ \cdots\ &&\ \cdots\ &&
\end{quantikz}}
\rightarrow^m
\scalebox{0.9}{
\begin{quantikz}[wire types = {b,b},classical gap=0.07cm, column sep = {1mm}]
&\gate{cs}\vqw{2} &[1mm]&\ \cdots\  \compgroup{3}{1}  &[2mm] \gate{cs}\vqw{2} &\\
& &&\ \cdots\ &  & \\
\lstick{$a_k$}&\targ{}&&\ \cdots\ &\targ{}&
\end{quantikz}}
\]
\begin{itemize}
\item \pn{pears}\emph{ definition}: the procedure statement $\ST^\pn{pears}$ is replaced by its definition. In the recursive case (input size larger than 2) the rule is:
\end{itemize}
\[
\scalebox{0.9}{
\begin{quantikz}[wire types = {b,q,q,q,b},classical gap=0.07cm, column sep = {1mm},row sep = {4mm,between origins}]
&&\ \cdots\  \compgroup{5}{3} & \gate{cs}\vqw{1} &\ \cdots\ & & \\[1mm]
&&\ \cdots\ &\gate[4]{}&\ \cdots\ &&\\
&&\ \cdots\ &&\ \cdots\ &&\\
&&\ \cdots\ &&\ \cdots\ &&\\
& &\ \cdots\ &  &\ \cdots\  & &
\end{quantikz}}
\rightarrow^d
\scalebox{0.9}{
\begin{quantikz}[wire types = {b,q,q,q,b},classical gap=0.07cm, column sep = {1mm},row sep = {4mm,between origins}]
&&\ \cdots\  \compgroup{5}{4} & \gate{cs}\vqw{1} & \gate{cs}\vqw{1} &\ \cdots\ & & \\[1mm]
&&\ \cdots\ &\octrl{1}&\ctrl{1}&\ \cdots\ &&\\
&&\ \cdots\ &\octrl{2}&\ctrl{1}&\ \cdots\ &&\\
&&\ \cdots\ &&\gate[2,style={fill=gray!50}]{}&\ \cdots\ &&\\
& &\ \cdots\ &\gate[style={fill=gray!50}]{} & &\ \cdots\  & &
\end{quantikz}}
\]

The compilation terminates once there are no more gates inside the dashed rectangle. Each instance of \pn{pears} on input size $k$ will have a corresponding ancilla $a_k$ for anchoring and merging.

The validity of the transformations as applied in Figure~\ref{fig:pears-compilation} is given in~\cite{HPS25}.

\section{Comparison with other languages}

We compare the \pfoqcompiler{} with current compilation techniques for recursive quantum control in Table~\ref{table:compilation-strategies}. We use the notation $\Tilde{\mathcal{O}}(\cdot)$ and $\mathcal{O}^\text{poly}(\cdot)$ to indicate that we ignore polylogarithmic and polynomial factors in the input size, respectively. We compare different aspects of circuit complexity (width, depth and size) for different programming languages. Highlighted in blue are the cases where the complexity scales as the maximum of the two branches.

We separate the languages according to whether the strategy used is sequential (as in Figure~\ref{fig:sequential-strategy}) or parallel (Figure~\ref{fig:parallel-strategy}). Among the strategies considered, only merging attains a complexity bound on circuit size that scales as the maximum complexity between branches. This can be viewed as the optimal bound in the sense that it avoids branch sequentialization. In the case where programs are also weakly uniform, the complexity matches that of classical control, with no polynomial slowdown.

We empirically test the impact of the different compilation techniques on the scaling of depth, width and size of circuits in Appendix~\ref{app:comparison} for Proto-Quipper~\cite{QUIPPER}, Qrisp~\cite{QRISP}, Qunity~\cite{QUNITY}, and Tower~\cite{TOWER}. This is done by writing a program equivalent to \texttt{\color{gray}PAIRS} in the target language and compiling it. If the language allows for a quantum switch statement, this is used directly.

We observe that, for the \texttt{\color{gray}PAIRS}-equivalent programs, each language compiles the program into circuits of exponentially-increasing size (that is, total number of wires and gates).

\section{Conclusion}

We have presented \pfoqcompiler{}, the first quantum programming language compiler that exploits program structure to improve asymptotic bounds of quantum circuits. This is achieved by restricting the set of programs to those with enough structure that simplifications can be made that reduce the complexity of recursive calls in superposition, avoiding the exponential blow-up of branch sequentialization.

The restrictions define three main fragments: $\pfoq{}^\star$ which characterizes quantum polynomial time programs, $\lfoq{}^\star$ which characterizes quantum polylogarithmic time programs, and the $\unif^\star$ fragment where branch sequentialization can be avoided altogether, which is particularly interesting when taken as the intersection $\pfoq{}^\star \cap \unif{}^\star$.

The restrictions provided here are relaxations of the ones given in~\cite{HPS23, HPS25, FHPS25}, with the intended purpose of bridging the gap between implicit characterizations and standard programming techniques. These restrictions can be quickly checked by the compiler and feedback is given as to where a restriction is broken so that the user can more easily adapt their program.

\section*{Acknowledgments}

This work has been partially funded by the European Union through the MSCA SE project QCOMICAL (Grant Agreement ID: 101182520), by the French National Research Agency (ANR): project TaQC ANR-22-CE47-0012, within the framework of ``Plan France 2030'', under the research projects EPIQ ANR-22-PETQ-0007 and HQI-R\&D ANR-22-PNCQ-0002, and by the project BPI France I-Demo AeroQat, contract 202400143.

\newpage

\bibliographystyle{alphaurl}
\bibliography{references}

\appendix

\newpage

\section{Comparison with other languages (continuation)}\label{app:comparison}

\begin{figure*}
     \centering
     \includegraphics[scale=0.9]{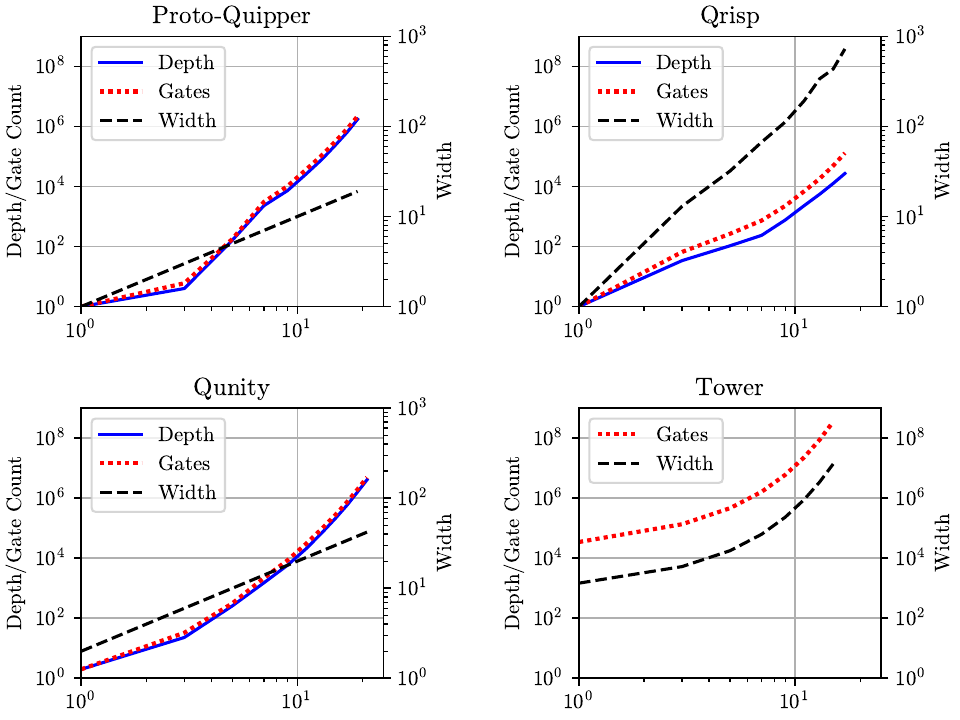}
\caption{Complexity metrics over different programming languages for a \texttt{\color{gray}PAIRS}-equivalent program. All plots are log-log and the $x$-axis is the input size.}
\label{fig:pairs comparison}
\end{figure*}

In this section, we consider the \texttt{\color{gray}PAIRS} program, implemented with a recursive quantum-control statement (or equivalent) in order to compare the scaling of different metrics of circuit complexity, namely number of qubits (including ancillae), circuit depth, and total number of gates. 
The programs implemented in the various languages, the materials and code needed to recreate Figure~\ref{fig:pairs comparison} are available in a Zenodo artifact~\cite{pfoqcompare}.\footnote{The figure was compiled using an Intel(R) Core(TM) Ultra 7 165H \@ 4.30 GHz $\times$ 16 cores processor and 32 GB of RAM using Ubuntu 24.04.4 LTS.}

\paragraph*{Qrisp.}
In Qrisp~\cite{QRISP}, we can make use of a quantum version of a switch statement, denoted \texttt{q\_switch}, which allows for the program to be structured in a similar way to the quantum conditionals in \foq{}.
This has the benefit of allowing different compilation techniques: \textrm{sequential}, \textrm{parallel}, or \textrm{tree} (which uses a balanced tree method to compile a circuit).

These different techniques compile a circuit that is efficient in either width or depth depending on the option given.
However, all of the techniques blow up the complexity of one of the metrics for the \texttt{\color{gray}PAIRS} program.
Using the multiplexor approach (\textrm{parallel}), the program needs to be modified to allow for extra ancillary qubits, which exponentially increases the width of the circuit.

\paragraph*{Proto-Quipper.}
Proto-Quipper~\cite{QUIPPER} allows for a \texttt{case} construction which adds quantum control to circuits.
Our approach to writing \texttt{\color{gray}PAIRS} in Proto-Quipper consists of recursively defining the \textbf{pairs} circuit using the \texttt{case} construction to control on the first two qubits.

The circuit is then compiled with a strategy akin to that of Figure~\ref{fig:sequential-strategy}, where the circuit depth increases exponentially with the input size to \textbf{pairs}.

\paragraph*{Qunity.}
Within Qunity, a \texttt{pmatch} statement, detailed in~\cite{QUNITY_compile}, is used to create a \texttt{\color{gray}PAIRS} program similar to the one written in \foq{}~(Figure~\ref{fig:pairs}).
Qunity also features \texttt{ctrl} and \texttt{match} statements to handle pattern matching on state vectors and density operators respectively. The choice of \texttt{pmatch} is justified since quantum branching can effectively be done on the pure states $\ket{00},\,\ket{01},\,\ket{10}$ and $\ket{11}$. \textbf{pairs} is recursively called on the $\ket{00}$ and $\ket{11}$ cases, and in the remaining cases nothing happens.

The Qunity compiler uses a tree-based compilation technique, which results in an exponential blow-up in the compiled circuit depth.

\paragraph*{Tower.}
The implementation in Tower~\cite{TOWER} is similar to that of Proto-Quipper, where we use appropriate conditional statements to recursively call \textbf{pairs} on the required states.

Unlike the other languages we consider, Tower requires a different data structure to be used, which is a quantum implementation of a list of booleans with pointers.
Due to this difference in data structure that is more general, the compilation of the program results in a very large circuit that still faces a blow-up in depth. The compilation of the different cases is performed sequentially (Figure~\ref{fig:sequential-strategy}).

As is the case for the previous languages, it is not particularly difficult to find a Tower program simulating \texttt{\color{gray}PAIRS} that avoids the exponential blow-up. In the case of Tower, the same trick used for avoiding branch sequentialization in the \textbf{search} program could be used here~\cite[Figure 24]{TOWER}. This would essentially consist of the merging strategy (Figure~\ref{fig:merging-strategy}). The point is mainly that a very similar program (take, for instance,  \texttt{\color{gray}PEARS} in Section~\ref{s:compilation-example}) would not admit such an immediate strategy.

\section{Proofs}
\label{app:proofs}

\staticanalysis*
\begin{proof}
Given a \foq{} program $\PR$, build the \emph{call graph} $\mathcal{G}$ for $\PR$, which consists of a weighted multigraph where the set of nodes is $\{\pn{proc$_i$}\}_{\pn{proc$_i$}\in\PR}$. Then, $\mathcal{G}$ contains the directed edge (\pn{proc$_i$}, \pn{proc$_j$}) if $\call $\pn{proc$_j$}[\ia](l$_1$,\dots,l$_k$)$\in \ST^\pn{proc$_i$}$ and the weight $w\in\{0,1,2\}$ of the edge is defined as 0 if the qubit lists $(l_1,\dots,l_k)$ do not reduce the available qubits, 1 if at least one $l_i$  reduces the amount of qubits by a constant amount, and 2 if at least one $l_i$ reduces the amount of qubits by half, where  the condition for the weight being 2 subsumes that of the weight being 1. Note that there may be more than one edge between two nodes, possibly of differing weight. $\mathcal{G}$ can be obtained from $\PR$ in linear time and is also bounded in number of edges and nodes by $\mathcal{O}(|\PR|)$.

The well-foundedness condition is then equivalent to the non-existence of a cycle in $\mathcal{G}$ consisting only of edges of weight 0; and the recursively halving condition is equivalent to the non-existence of a cycle consisting only of edges of weight 0 or 1. These subgraphs of $\mathcal{G}$ can be obtained in time linear in its size, and determining whether the subgraph contains a cycle can be done in linear time on its size.

Verifying the bounded width condition amounts to computing the width of each procedure. This can be done by computing the width of each statement appearing in the abstract syntax tree of the procedure statement, which can be done in linear time on its size. The total number of procedures in the program is also bounded by $\mathcal{O}(|\PR|)$, giving a final quadratic bound.

To check that a program is weakly uniform, we use the width information obtained from computing the bounded width condition to determine what procedure calls are made in branches that are orthogonal to the recursive branches. The condition can then be easily checked in linear time on each procedure statement.
\end{proof}

\wfstar*
\begin{proof}
In order to ensure that the semantics is preserved in the transformation, we explicitly state the behavior of each procedure on input size zero. Let $\mathcal{X}$ denote some sequence of integer variables. For each procedure \pn{proc}, we rewrite its procedure statement as follows:
\begin{center}
\begin{PFOQprogram}
 \tbt{decl }\proc[$\mathcal{X}$](q$_1$,\dots,q$_k$)\{\ST$^\proc$\}
\end{PFOQprogram}
\end{center}
becomes
\begin{center}
\scalebox{0.9}{
\begin{PFOQprogram}
 \tbt{decl }\proc[$\mathcal{X}$](q$_1$,\dots,q$_k$)\{\\
 \quad \tbt{if }(|q$_1$|>0 \tbt{\&\&}\,\dots\,\tbt{\&\&} |q$_k$|>0) \tbt{then} \{S$^\proc$\}\\
 \quad \tbt{else} \{ \tbt{skip}; \}\}
\end{PFOQprogram}}

\end{center}

Note that this transformation does not change the program's semantics. For simplicity, we consider that $\PR$ is already in this form at the start.

To obtain $\PR'$ from $\PR$, we again transform the procedure statements in a semantics-preserving way. Let $\pn{proc} \in \PR$ be a  procedure. Then, for every $\pn{proc'} \in \PR$ such that $\proc\sim \pn{proc'}$ and $\call \pn{proc'} \in \ST^\proc$ does \emph{not} reduce the size of any input qubit list, then we replace $\call \pn{proc'}$ with $\ST^{\pn{proc'}}$. We repeat this step until all procedure calls in $\ST^\proc$ reduce the size of some qubit lists.

The zero size condition we explicitly included in the procedure statement ensures that this transformation preserves the procedure semantics. Once the transformation finishes, for all procedures, we obtain the new program $\PR'$ that is strongly well-founded, which has the same behavior as $\PR$.

If $\PR$ is weakly well-founded, then this transformation terminates. This can be seen from the fact that, if the procedure does not terminate, there is a sequence
\[\tbt{call }\pn{proc$_{i+1}$}\in\ST^{\pn{proc$_{i}$}}\]
such that $\pn{proc$_{k}$} = \pn{proc$_{0}$}$ and none of the procedure calls reduce the number of input qubits.
\end{proof}

\characterization*
\begin{proof}
In \cite[Theorem 3]{HPS23} it is shown that the set of programs $\wf{} \cap \wi_{\leq 1}$ (where the well-foundedness restriction is strong) is sound and complete for polynomial time. Completeness is then trivial since $\wf{} \cap \wi_{\leq 1}\ \subsetneq\ \pfoq{}^\star$. Soundness can be demonstrated from the fact that the transformation of $\wf{}^\star$ programs into $\wf{}$ (proof of Theorem~\ref{thm:wfstar}) does not change the program's execution time.
\end{proof}

\unifcharacterization*
\begin{proof}
Soundness is trivial since $\pfoq{}^\star$ is sound for quantum polynomial time (Theorem~\ref{thm:characterization}). Completeness is given from the fact that $\pfoq{}\cap\unif{}$ is complete for quantum polynomial time~\cite{HPS25}.
\end{proof}

\weakunifcirc*
\begin{proof}
This is proven in~\cite[Theorem 9]{HPS25} for the $\pfoq{}\cap \unif{}$ fragment, i.e. where all the restrictions are strong. In the weak version, we consider similar arguments made more precise.

The proof is done by checking the circuit complexity of each possible statement, and comparing it with the \emph{time} complexity in the operational semantics. The non-trivial case of this theorem is the  \tbt{qcase} statement, where the circuit complexity should scale as the maximum complexity of the branches. In the context of some recursive procedure $\proc{}$, we consider the cases where the two branches are recursive, and when only one is recursive.

The weak restriction $\pfoq{}^\star$ combined with conditions \ref{unif-i} and \ref{unif-ii} of weak uniformity (Definition~\ref{def:weak-uniformity}) ensure that a recursive procedure $\proc\in\PR$ on input size $n$ will generate at most $O(n)$ procedure calls to procedures $\pn{proc'}$ such that $\proc\sim \pn{proc'}$ (up to the compatibility criteria that allow for merging). This guarantees that at each level of recursion at most $n$ unique procedure calls are compiled, which is our desired bound -- a similar argument can be made for when \pn{proc} terminates in polylogarithmic time. 

We consider now the case of a \tbt{qcase} with a non-recursive branch containing calls to procedures $\pn{proc''}$ such that $\proc \geq \pn{proc''}$ and $\proc\not\sim \pn{proc''}$. By condition~\ref{unif-iii} these calls should also be similarly reducing and can be merged without overhead. Therefore, we obtain the desired bound on the complexity of the \tbt{qcase}, which concludes the proof.
\end{proof}

\end{document}